\documentclass[10pt,a4paper]{article}
\usepackage[utf8]{inputenc}
\usepackage[english]{babel}
\usepackage{amsmath}
\usepackage{amsfonts}
\usepackage{amssymb}
\usepackage{graphicx}
\usepackage{color}
\usepackage[left=2cm,right=2cm,top=2cm,bottom=2cm]{geometry}
\usepackage{multirow}
\usepackage{cite}
\usepackage{authblk}
\usepackage{hyperref}

\author[1,2]{Barnabás Pórfy}
\author[1]{Máté Csanád}
\affil[1]{Department of Atomic Physics, ELTE E\"otv\"os Lor\'and University, Budapest, Hungary}
\affil[2]{HUN-REN Wigner Research Centre for Physics, Budapest, Hungary}
\title{Geometry of particle emission in UrQMD Ar+Sc collisions at SPS energies}
\begin{document}
\maketitle

\begin{abstract}
Over the past few decades, progress in femtoscopy has been driven by the interplay between experimental measurements and theoretical calculations. Measurements provide data to support the theory, while theoretical predictions guide new measurements. In the recent decade, several experiments have confirmed that the two-particle pion-emitting source is well described by Lévy alpha-stable distributions. To enable theoretical interpretation, phenomenological simulations have been done at RHIC and LHC energies, in large systems such as Au+Au or Pb+Pb, using various available heavy-ion collision models. However, such simulations have not been done in intermediate systems. In this paper, we investigate three-dimensional two-pion pair source distributions from  $^{40}$Ar+$^{45}$Sc central collisions at SPS energies, generated with the Ultra-Relativistic Quantum Molecular Dynamics Monte-Carlo event generator. Supplemented by a validation of the simulated hadronic spectra, we find that the pair source can be described with Lévy-stable distributions. We subsequently interpret the physical meaning of the extracted Lévy parameters corresponding to the spatial scale, shape, and strength of the source. Our results form a baseline for future experimental measurements in intermediate systems.

\end{abstract}

\section{Introduction}\label{s:intro}
The general goal of high-energy heavy-ion experiments is to explore the properties of the strongly interacting matter created in heavy-ion collisions. A key objective is to understand the space–time geometry of the particle-emitting source. One approach to this is femtoscopy, i.e., the analysis of quantum-statistical correlations of particles produced in such collisions~\cite{Goldhaber:1959mj, Goldhaber:1960sf}. Experimental results based on the so-called imaging technique~\cite{Brown:1997ku} around the turn of the millennium~\cite{PHENIX:2006nml}, together with more recent measurements~\cite{PHENIX:2017ino,NA61SHINE:2023qzr,CMS:2023xyd,Kincses:2024sin,PHENIX:2024vjp}, indicate that the commonly assumed Gaussian shape of the source~\cite{Csorgo:1994fg, Akkelin:1995gh} does not provide an adequate description. Instead, the two-pion source function exhibits a power-law tail, and a Lévy-type parametrization offers a statistically robust description of the measured femtoscopic correlation functions across collision energies.~\cite{Csanad:2024hva}

Several physical mechanisms may contribute to the appearance of Lévy distributions in high-energy heavy-ion collisions. Short-lived resonance decays, as well as elastic and inelastic processes occur over many timescales during and shortly after the freeze-out. As demonstrated in Ref.~\cite{Kincses:2024lnv}, these lead to the phenomenon of Lévy walk, which in turn results in Lévy distributed sources~\cite{Csorgo:2003uv,Csanad:2007fr}. In terms of the core-halo picture~\cite{Csorgo:1994in}, the source core component can then be well described by a single Lévy distribution (as opposed to a Gaussian with an exponential or power-law tail)~\cite{Kincses:2025izu}. It was also demonstrated in event-by-event and three-dimensional analyses that directional and event averaging in experimental analyses does not influence this behavior~\cite{Kincses:2022eqq, Korodi:2022ohn, Kincses:2025izu}. Additional possible sources of Lévy-type emission patterns include jet fragmentation~\cite{Csorgo:2004sr} and critical dynamics near the QCD critical point~\cite{Csorgo:2005it}. To investigate these underlying processes, phenomenological studies with event generators are particularly valuable. An available option is to use the Ultra-Relativistic Quantum Molecular Dynamics (UrQMD) framework~\cite{Bass:1998ca, Bleicher:1999xi}. The clear advantage of using such event generators is the direct reconstruction of the source function; therefore, hidden properties of experimental data may be uncovered by investigating simulated events. 

In this paper, we present a detailed investigation of the two-particle source function using like-charged pions in argon–scandium ($^{40}$Ar+$^{45}$Sc) collision data at the available energy range of the NA61/SHINE experiment~\cite{Abgrall:2014xwa} ($\sqrt{s_\text{NN}} \approx $5-17 GeV, or 13-150A GeV/$c$ beam momentum) generated by the UrQMD model.
The reconstructed two-particle source functions are fitted with three-dimensional Lévy distributions~\cite{LevyCode}, similarly to Ref.~\cite{Kincses:2025izu}. The source parameters are studied as functions of pair transverse mass and collision energy.

The paper is structured as follows: In Sec.~\ref{s:source}, the properties of the two-particle source function are detailed alongside the Lévy approach and the event generator, and the following Sec.~\ref{s:analysis} gives the details on the measurement and analysis itself. The results of Lévy parameters as a function of transverse mass as well as a function of collision energy are detailed in Sec.~\ref{s:res}. Finally, the conclusions are presented in Sec.~\ref{s:con}.

\section{Two-particle source function}\label{s:source}
The investigation of quantum-statistical correlation functions is one of the most important methods in heavy-ion physics. Femtoscopic correlations provide information about the space-time geometry of the particle-emitting source~\cite{Lisa:2005dd, Csorgo:1999sj}. The connection between the two-particle momentum correlations ($C_2$) and the pair source distribution ($D$) is as follows:
\begin{equation}\label{eq:corrfunc}
	C_2(q,K) = \int d^4r D(r,K)|\Psi^{(2)}_q(r)|^2,
\end{equation}
where $q = p_1 - p_2$ is the momentum difference of the pair, $K = (p_1 + p_2)/2$ signifies the average pair momentum and $r = x_1 - x_2$ denotes the four vector distance of the pair, while $\Psi^{(2)}_q(r)$ is the pair wave function. Thus, through measuring the correlation function, femtoscopy maps the pair source distribution and the pair wave function. The pair source is the autocorrelation of single-particle source functions ($S$) and is introduced as
\begin{equation}\label{eq:pairsource}
	D(r,K) = \int d^4\sigma S(\sigma + r/2,K)S(\sigma - r/2,K),
\end{equation}
where $\sigma = (x_1 + x_2)/2$ is the pair center of mass. It is important to note that in both Eq.~\eqref{eq:corrfunc} and Eq.~\eqref{eq:pairsource} the dependence on $K$ is assumed to be weaker~\cite{Lisa:2005dd} and is often appears only through the dependence of parameters on it. While for the shape of the source, historically a Gaussian was assumed~\cite{Akkelin:1995gh}, a more general approach is a three-dimensional Lévy-stable distribution~\cite{Csorgo:2003uv}
\begin{equation}
	\mathcal{L}(\alpha, R^2, \vec{r}) = \frac{1}{(2\pi)^3} 
\int d^3\vec{\zeta} \, e^{i \vec{\zeta} \cdot \vec{r}} \, e^{-\frac{1}{2} \left| \vec{\zeta}^T R^2 \vec{\zeta} \right|^{\alpha/2}},\label{e:Levydef}
\end{equation}
where the parameters of this distribution are $\alpha$, the Lévy stability index, and the $R^2$ matrix of Lévy scale parameters or radii, often assumed to be diagonal, i.e., $\mathrm{diag}\left(R_{\mathrm{out}}^2, R_{\mathrm{side}}^2, R_{\mathrm{long}}^2\right)$ in the Bertsch--Pratt coordinate system~\cite{Bertsch:1988db, Pratt:1990zq}. Furthermore, $\vec r$ is the vector of spatial coordinates, and the vector $\vec \zeta$ represents the integration variable. In the case of $\alpha=2$, one recovers the Gaussian distribution, while $\alpha = 1$ is equivalent to the Cauchy distribution, and for $\alpha < 2$, the Lévy distribution exhibits a power-law tail. The defining ``stability'' property of these distributions is that they retain the same shape parameter under convolution. Consequently, if single particle sources are Lévy distributed, the pair source distribution is also Lévy shaped with the same $\alpha$ value, and Lévy radii are scaled by a factor of $2^{1/\alpha}$~\cite{PHENIX:2017ino}.

In experimental heavy-ion physics, one has no direct access to the freeze-out distribution. However, femtoscopic and imaging techniques do allow one to constrain this distribution via measurements of momentum correlations. On the other hand, in event generators, one can obtain the freeze-out coordinates corresponding to the last point of interaction of each produced particle, from which one can calculate the freeze-out distance distributions of the given event generator. One such generator is the UrQMD framework~\cite{Bass:1998ca,Bleicher:1999xi}, detailed in the following part. 

\section{Analysis details}\label{s:analysis}
\subsection{UrQMD}
The UrQMD event generator is a microscopic transport model widely used to simulate heavy-ion collisions over a broad range of energies. The model is well suited to be applied in the 2-160\textit{A} GeV beam energy region, in line with the energy range available at the NA61/SHINE experiment~\cite{Abgrall:2014xwa}. We employ the UrQMD model (v3.4) in cascade mode as the main event generator.
For this study, UrQMD was slightly altered to include the decay of $\eta$ mesons, based on the work detailed in~\cite{Uzhinsky:2013ata}, in order to improve the connection to real data, as the $\eta$ meson is the only one among the variety of particles produced in UrQMD that may have an influence on the source distribution, see Fig. 1 of Ref~\cite{Kincses:2025izu}. However, this only affects the strength of the extracted source.

Simulations were performed for beam momenta of 13–150\textit{A} GeV/$c$ in the $^{40}$Ar+$^{45}$Sc collision system, corresponding to center of mass energies per nucleon pair of $\sqrt{s_\text{NN}} \approx $ 5-17 GeV. For each energy, 10,000 events were generated in the 0-10\% centrality range with time evolution duration set to 10,000 fm/\textit{c}. To align with the experimental analyses, centrality selection was defined by assessing the total energy in the kinematical range of the Projectile Spectator Detector (PSD) used for centrality selection in the NA61/SHINE experiment, detailed in~\cite{Abgrall:2014xwa}. Events with the smallest amount of PSD-range energy ($E_{\rm PSD}$) can be considered most central. The $E_{\rm PSD}$ limit for 0-10\% centrality range was then determined based on Minimum Bias UrQMD simulations.

A comparison of UrQMD results of transverse momentum and rapidity distributions to NA61/SHINE data from 13 to 150$A$ GeV/$c$ Ar+Sc collisions~\cite{NA61SHINE:2023epu} is given in Appendix~\ref{a:spectra}. These indicate that transverse momentum spectra and rapidity distributions of pions from UrQMD describe data well. Mean multiplicities are slightly underestimated for pions and an even stronger discrepancy is seen for kaons, compared to available data. Kaon $p_T$ spectra shapes are well captured, but overall magnitudes are underestimated at some energies, similarly to d$n$/d$y$ at around midrapidity.  These are in line with the overall multiplicity difference between data and UrQMD for kaons. Finally, both $p_T$ distributions and d$n$/d$y$ data for protons and antiprotons are well described by UrQMD. These indicate that UrQMD performs comparably to PHSD, SMASH, and EPOS1.99~\cite{NA61SHINE:2023epu}.

\subsection{Pair source function}
The choice of frame is essential for femtoscopic studies, and two options are often chosen: the pair center of mass system (PCMS) and the longitudinally comoving system (LCMS). In case of the PCMS, the total three-momentum vanishes. On the other hand, in the LCMS frame, for mid-rapidity measurements the cross-terms will vanish~\cite{Chapman:1994yv, PHENIX:2004yan}, rendering the source nearly spherically symmetric~\cite{PHENIX:2017ino}. In the LCMS, the longitudinal component (beam direction) of the pair average momentum is zero by definition. Additionally, one often uses the Bertsch-Pratt coordinate system~\cite{Pratt:1986ev,Bertsch:1988db}, where ``long'' corresponds to the aforementioned longitudinal direction, ``out'' direction is the direction of the average pair momentum perpendicular to ``long''. Finally, ``side'' is perpendicular to both directions.
To access the pair source function, we follow the definition of Refs.~\cite{Kincses:2024lnv,Kincses:2025iaf,Kincses:2025izu} for the three-dimensional spatial distance vector (in $c=1$ units):
\begin{align}
    \vec{\rho} &= \vec{r} - \frac{\vec{K}}{K_0} t.
\end{align}
Its components in the LCMS can be expressed from lab-frame coordinates and momenta as~\cite{Kincses:2024lnv,Kincses:2025iaf,Kincses:2025izu,Huang:2025edi}
\begin{align}
	\rho_\text{out}^\text{LCMS} &= r_x \cos \varphi + r_y \sin \varphi - \frac{K_\text{T}}{K^2_0 - K_z^2} \left(K_0t - K_z r_z\right) \\
	\rho_\text{side}^\text{LCMS} &= -r_x \sin \varphi + r_y \cos \varphi \\
	\rho_\text{long}^\text{LCMS} &= \frac{K_0 r_z - K_z t}{\sqrt{K^2_0 - K_z^2}},
\end{align}
where $\cos \varphi = K_x/K_\text{T}, \sin \varphi = K_y/K_\text{T}$, and $K_\text{T}^2 = K_x^2 + K_y^2$.

In this analysis, we calculate the one-dimensional projections of the three-dimensional $D(\vec \rho)$ distribution: 
\begin{align*}
D\left(\rho_\text{out}^\text{LCMS}\right),\qquad
D\left(\rho_\text{side}^\text{LCMS}\right),\qquad
D\left(\rho_\text{long}^\text{LCMS}\right),
\end{align*}
and fit them with the projections of Lévy distributions, as defined in Eq.~\eqref{e:Levydef}. Unlike the gold-gold or lead-lead collision analyzed Refs.~\cite{Korodi:2022ohn,Kincses:2022eqq}, single Ar+Sc events lack sufficient statistics for this procedure, hence, similarly to Refs.~\cite{Kincses:2025izu, Huang:2025edi}, we sum the $D$ distributions of several events. The fit parameters converge as the number of summed events ($N_{\rm events}$) increases, thus we can chose a sufficiently large $N_{\rm events}$ value, and then average the results from such blocks of events. As a default, we use $N_{\rm events}=2500$ for $p_{beam} =  150, 75$ and $40$\textit{A} GeV/\textit{c}, and 5000 for $p_{beam} = 30, 19$ and 13\textit{A} GeV/\textit{c}. To create conditions similar to the experimental analysis, several kinematic cuts are also applied: pions with single-track $y$ $\pm$ 1 (1.5 for 150\textit{A} GeV/$c$) around midrapidity, and with transverse momentum $p_\text{T} < 1.5$, are selected. Furthermore, to replicate the effect of the limited fit range in the experimental momentum correlation functions, we select pairs with a limited momentum difference: a $q_\text{LCMS} < \sqrt{m_\text{T}\cdot 150\;\textrm{MeV}/c^2}$ cut is applied. We then investigate the $D(\rho)$ distributions in eight $K_\text{T}$ classes ranging from 0 to 1 GeV/\textit{c}, similarly to the experimental analysis of Ref.~\cite{Porfy:2024jsu}.

\begin{figure}
    \centering
    \includegraphics[width=1\linewidth]{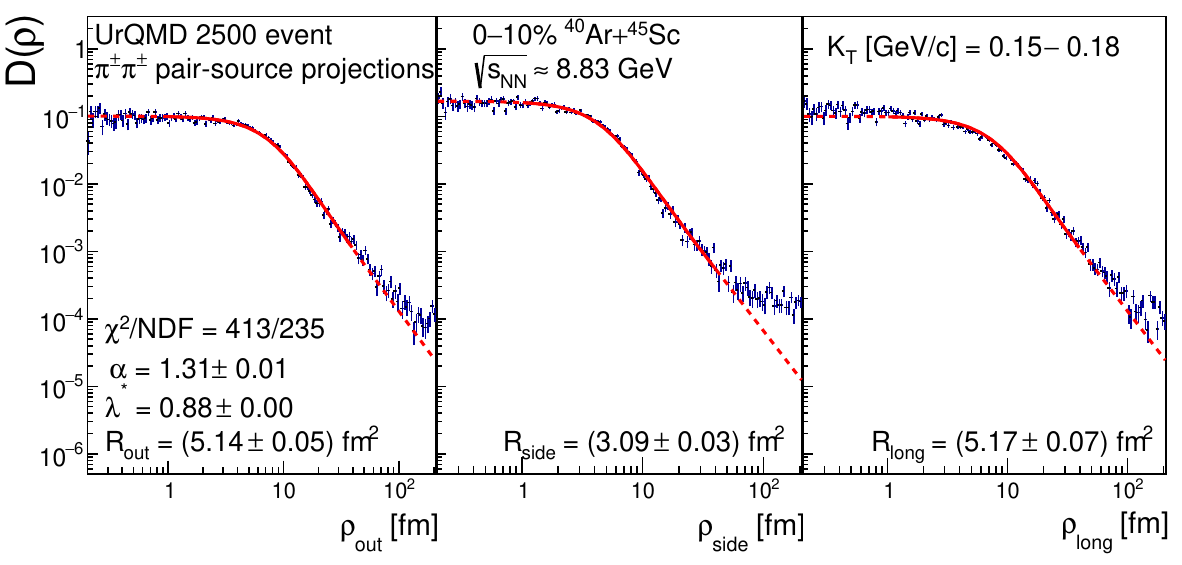}
    \caption{Example fit (red line) to projected $D(\rho)$ distributions (blue points) in Ar+Sc 0-10\% centrality at 40\textit{A} GeV/$c$ in $K_\text{T}$ range [0.15-0.18] GeV/$c$ with 2500 events merged. The fit is marked with a solid line, while the dashed line is an extrapolation.}
    \label{fig:fit_example}
\end{figure}
\subsection{Systematic Uncertainties}

In the analysis, we carefully consider uncertainties as well. The statistical uncertainties depend on the number of events considered and are negligible in our case. Systematic uncertainties, on the other hand, are of high importance. To assess these, we examined several different settings to understand their impact on the data. The settings are summarized in the Table~\ref{tab:systematic_sources}. 
\begin{table*}[h!]
\centering
\begin{tabular}{cccc}
\hline
Energy &
  $N_\text{events}$ &
  $Q^\text{max}_\text{LCMS}$ {[}MeV/$c${]} &
  Fit range {[}fm{]} \\ \hline
150\textit{A} GeV/$c$ &
  \multirow{3}{*}{\begin{tabular}[c]{@{}c@{}}standard: 2500\\ strict: 1000\\ loose: 5000\end{tabular}} &
  \multirow{6}{*}{\begin{tabular}[c]{@{}c@{}}standard: $\sqrt{150 \text{ MeV/}c^2 \cdot m_\text{T}}$\\ strict: $\sqrt{100\text{ MeV/}c^2 \cdot m_\text{T}}$\\ loose: $\sqrt{250\text{ MeV/}c^2\cdot m_\text{T}}$\end{tabular}} &
  \multirow{6}{*}{\begin{tabular}[c]{@{}c@{}}standard: 1-40\\ strict: 1-30\\ loose: 1-50\end{tabular}} \\ \cline{1-1}
75\textit{A} GeV/$c$ &
   &
   &
   \\ \cline{1-1}
40\textit{A} GeV/$c$ &
   &
   &
   \\ \cline{1-2}
30\textit{A} GeV/$c$ &
  \multirow{3}{*}{\begin{tabular}[c]{@{}c@{}}standard: 5000\\ strict: 2500\\ loose: 10000\end{tabular}} &
   &
   \\ \cline{1-1}
19\textit{A} GeV/$c$ &
   &
   &
   \\ \cline{1-1}
13\textit{A} GeV/$c$ &
   &
   &
   \\ \hline
\end{tabular}
\caption{Systematic uncertainty sources, their different settings, and values for all energies.}
\label{tab:systematic_sources}
\end{table*}
The total systematic uncertainty is then calculated, based on the default value ($P^{\rm def}$) of a parameter, its varied value ($P^{\rm var}_i$) for the $i$th source (among the $n=3$ sources: $N_\text{events}$, $Q^\text{max}_\text{LCMS}$, fit range), separately for the positive and negative uncertainties ($j=+,-$):
\begin{align}
    \sigma^2_{i,j}[P] &=(P^\text{def}-P^{\rm var}_{i,j})^2\\
    \sigma^\text{tot}_j[P] &= \sqrt{\sum^n_{i=1} \sigma^2_{i,j}[P]}.
\end{align}
This total uncertainty $\sigma^\text{tot}_\pm$ is subsequently shown on the results as an uncertainty band. The averaged systematic uncertainty for three energies is summarized in Table~\ref{tab:sys_sum}.
\begin{table*}[h!]
\small
\centering
\begin{tabular}{cccccccc}
\hline
$p_{beam}$ [\textit{A} GeV/$c$] & Source & $\alpha$ [\%] & $\lambda$ [\%] & $\bar{R}$ [\%] & $R_\text{out}$ [\%] & $R_\text{side}$ [\%] & $R_\text{long}$ [\%] \\ \hline
\multirow{4}{*}{150} & $Q^\text{max}_\text{LCMS}$ & 1.84 & 0.91 & 1.05 & 1.12 & 0.82 & 1.07 \\
                     & Fit range                  & 0.27 & 0.11 & 0.57 & 0.76 & 0.42 & 0.99 \\
                     & N$_\text{events}$          & 0.03 & 0.02 & 0.02 & 0.02 & 0.01 & 0.03 \\
                     & Total uncertainty          & 1.90 & 0.93 & 1.42 & 1.54 & 1.12 & 1.88 \\ \hline
\multirow{4}{*}{40}  & $Q^\text{max}_\text{LCMS}$ & 0.86 & 0.28 & 0.15 & 0.14 & 0.06 & 0.20 \\
                     & Fit range                  & 0.29 & 0.17 & 0.54 & 0.82 & 0.28 & 0.98 \\
                     & N$_\text{events}$          & 0.08 & 0.02 & 0.01 & 0.01 & 0.02 & 0.02 \\
                     & Total uncertainty          & 0.98 & 0.35 & 0.58 & 0.85 & 0.29 & 1.03 \\ \hline
\multirow{4}{*}{13}  & $Q^\text{max}_\text{LCMS}$ & 0.57 & 0.15 & 0.06 & 0.05 & 0.06 & 0.11 \\
                     & Fit range                  & 0.32 & 0.19 & 0.63 & 0.64 & 0.43 & 1.16 \\
                     & N$_\text{events}$          & 0.15 & 0.05 & 0.02 & 0.01 & 0.01 & 0.04 \\
                     & Total uncertainty          & 0.78 & 0.30 & 0.64 & 0.65 & 0.45 & 1.18 \\ \hline
\end{tabular}
\caption{Average relative systematic uncertainties over all $K_\text{T}$ intervals for selected energies.}
\label{tab:sys_sum}
\end{table*}

\section{Results}\label{s:res}

\begin{figure*}
     \includegraphics[width=.33\textwidth,trim=8 4 21 4,clip]{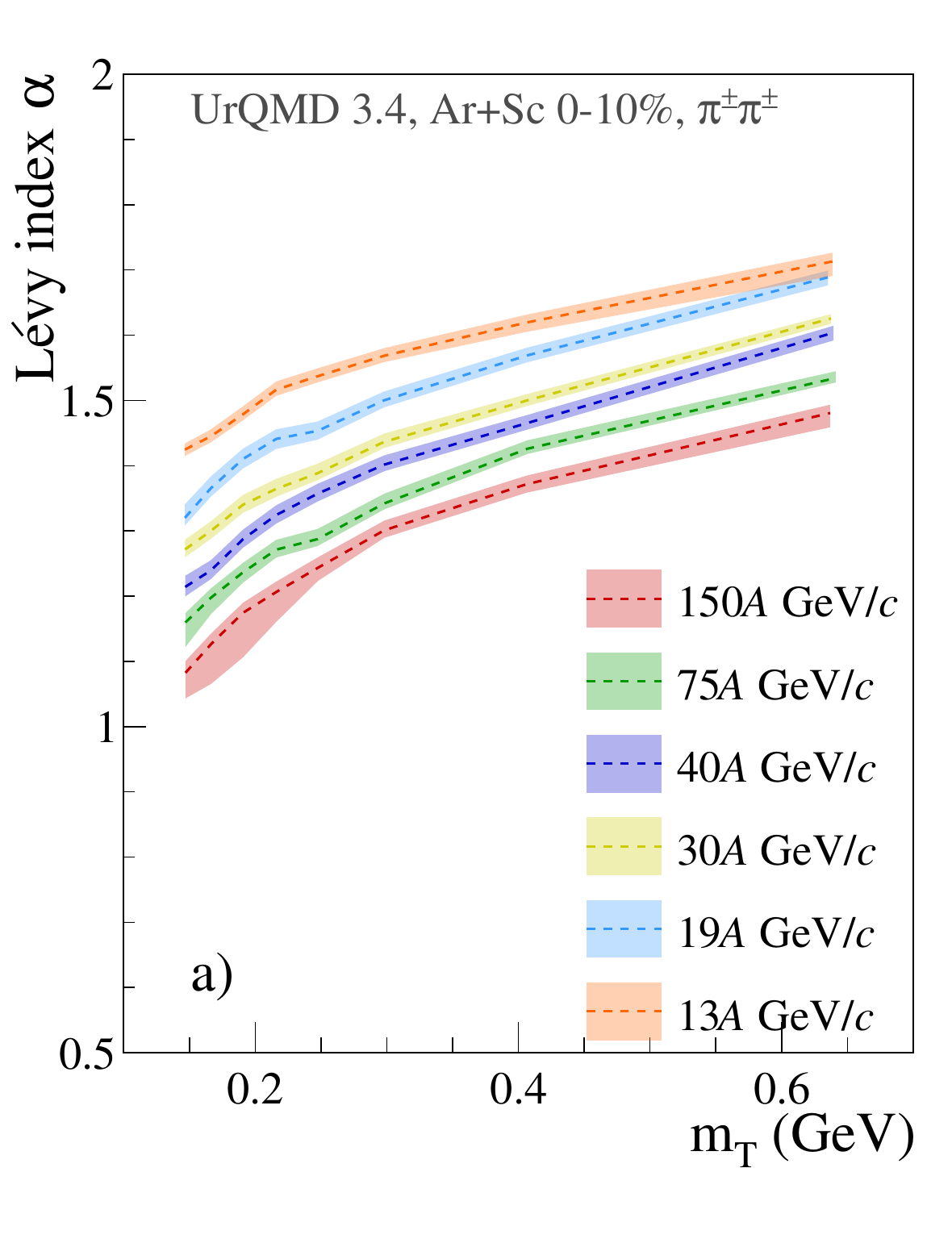}
     \includegraphics[width=.33\textwidth,trim=8 4 21 4,clip]{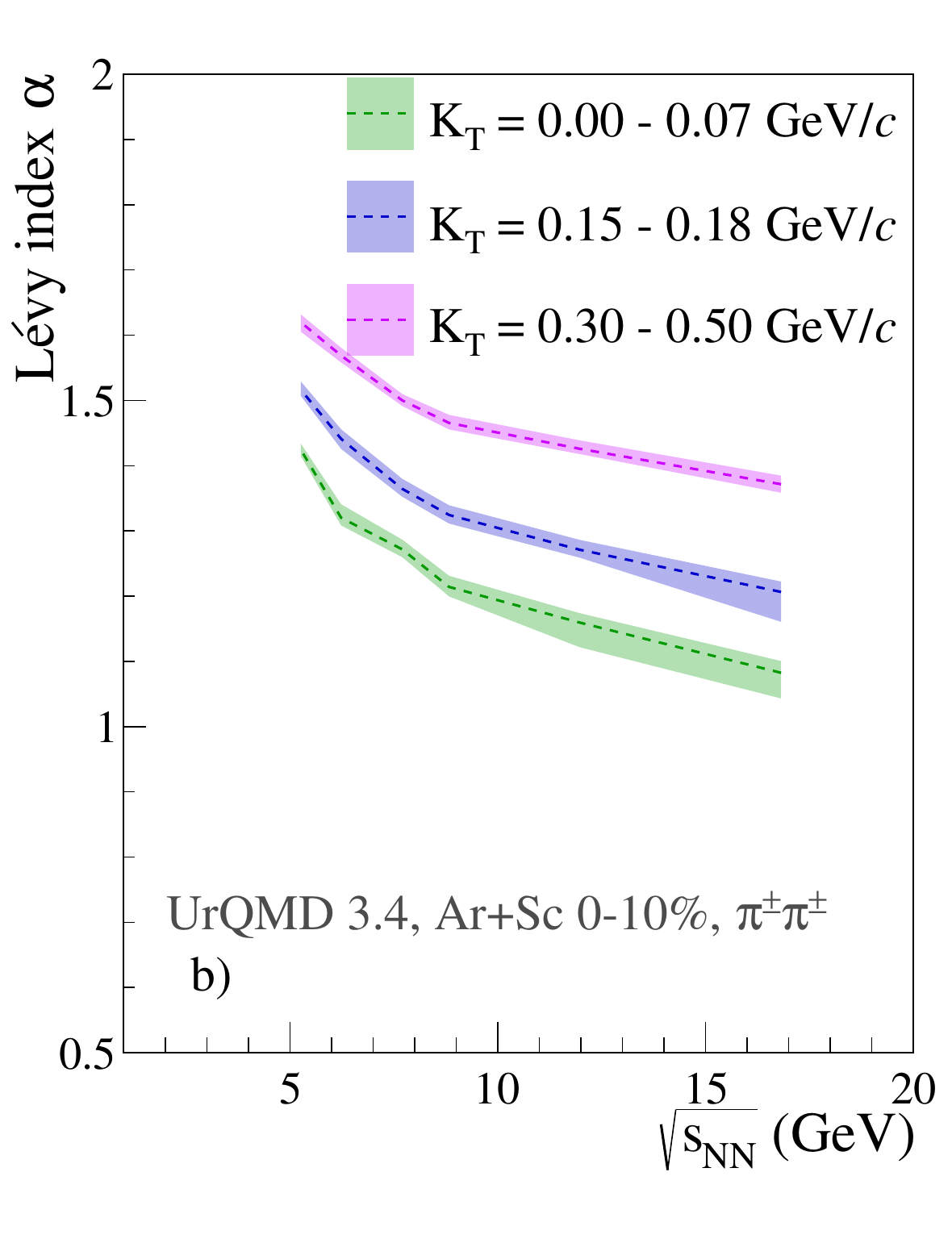}
     \includegraphics[width=.33\textwidth,trim=6 4 21 4,clip]{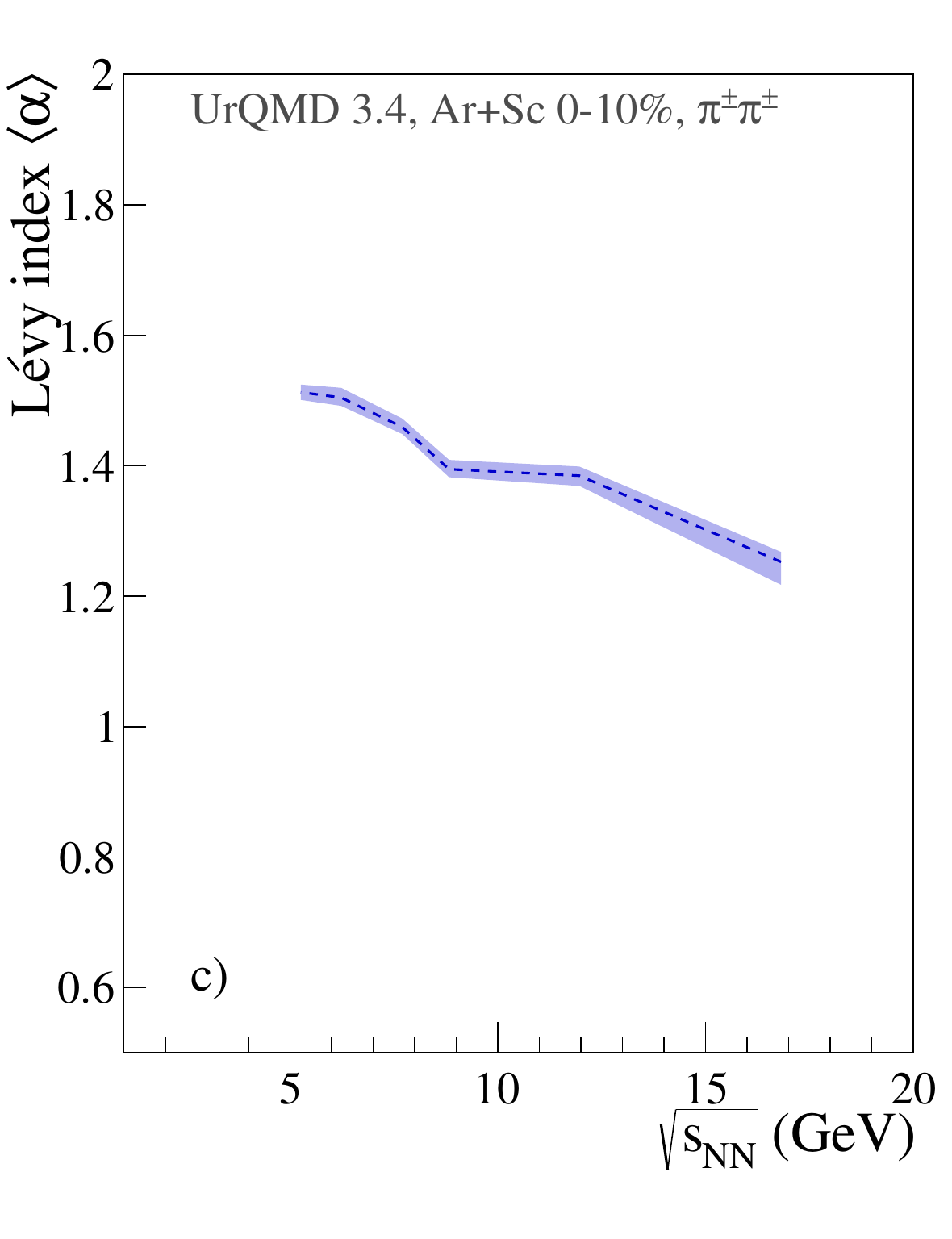}
     \caption{The Lévy index parameter $\alpha$, for 0--10\% central Ar+Sc. For all transverse mass $m_\text{T}$ a),  for selected transverse mass $m_\text{T}$ intervals b), and the average of all transverse mass $m_\text{T}$ intervals at the given energies c). The given colored band shows systematic uncertainty.}
     \label{fig:As}
\end{figure*}

Let us now turn towards the extracted fit parameters. Fig.~\ref{fig:As}a shows the Lévy index $\alpha$ as a function of $m_\text{T}$ for all simulated beam momenta (collision energies). It is apparent that $\alpha$ has a moderate dependence on $m_\text{T}$: a slight increase with $m_\text{T}$ is observable for all collision energies. This can be explained within the UrQMD framework by fewer resonance decays producing pions at larger momentum scales, in line with the increase of $\lambda$ towards larger momenta, discussed below. Furthermore, a point-by-point decrease of $\alpha({m_\text{T}})$ with collision energy is visible. To investigate this, $\alpha$ versus $\sqrt{s_\text{NN}}$ is shown in Fig.~\ref{fig:As}b in three selected (low, medium, and high) $m_\text{T}$ intervals. The collision energy dependence is qualitatively similar at all transverse momenta, but clear quantitative differences are observable. In particular, at the lowest $m_\text{T}$, $\alpha$ versus $\sqrt{s_\text{NN}}$ rapidly decreases to a value around unity and appears to flatten out, while for the highest $m_{\rm T}$ a smooth and continuous decrease is observable. To further investigate the overall trends, the $m_{\rm T}$-averaged $\langle \alpha \rangle$ is shown as a function of collision energy ($\sqrt{s_\text{NN}}$) in Fig.~\ref{fig:As}c, where the decreasing trend with increasing energy is observable. Such a trend implies the increase of the importance of the Lévy walk at higher energies in UrQMD. This is in contrast to a naive expectation of increased densities at higher energies leading to more conventional (as opposed to anomalous) diffusion. On the other hand, at higher energies, the production of heavier resonances increases, thus more short-lived decay pions can be expected, pronouncing the power-law tail and decreasing $\alpha$. In line with this, in the more diverse environment at higher energies, more processes (coalescence and inelastic scatterings) are possible, which, in the end, lead to a stronger Lévy walk. However, a clear understanding of these competing phenomena would require comparison of results in systems of different sizes, as well as to data.
The observed clear decrease is furthermore in contrast to preliminary NA61/SHINE data~\cite{Porfy:2024jsu} where an opposite trend was seen. This could be attributed to the onset of fireball in the experimental data, as UrQMD contains only the hadronic phase, driving the obtained $\alpha$ towards smaller values, as discussed in Ref.~\cite{Kincses:2024lnv}. In addition, an approximate minimum at around $\sqrt{s_\text{NN}} \approx 6-8$ GeV was seen in the preliminary data~\cite{Porfy:2024jsu}, which is absent in the simulations. Such a minimum was predicted~\cite{Csorgo:2005it} to appear in the vicinity of the critical point, a feature absent in UrQMD simulations. Overall, the discrepancy between model and data could be attributed to a lack of a hydrodynamic medium or criticality in UrQMD. Hadronic physics itself could also be responsible for this discrepancy, given the imperfect description of hadronic yields in UrQMD (see Appendix~\ref{a:spectra} and Section~\ref{s:analysis}).

\begin{figure}
     \centering
     \includegraphics[width=.5\textwidth, trim=2 4 21 4,clip]{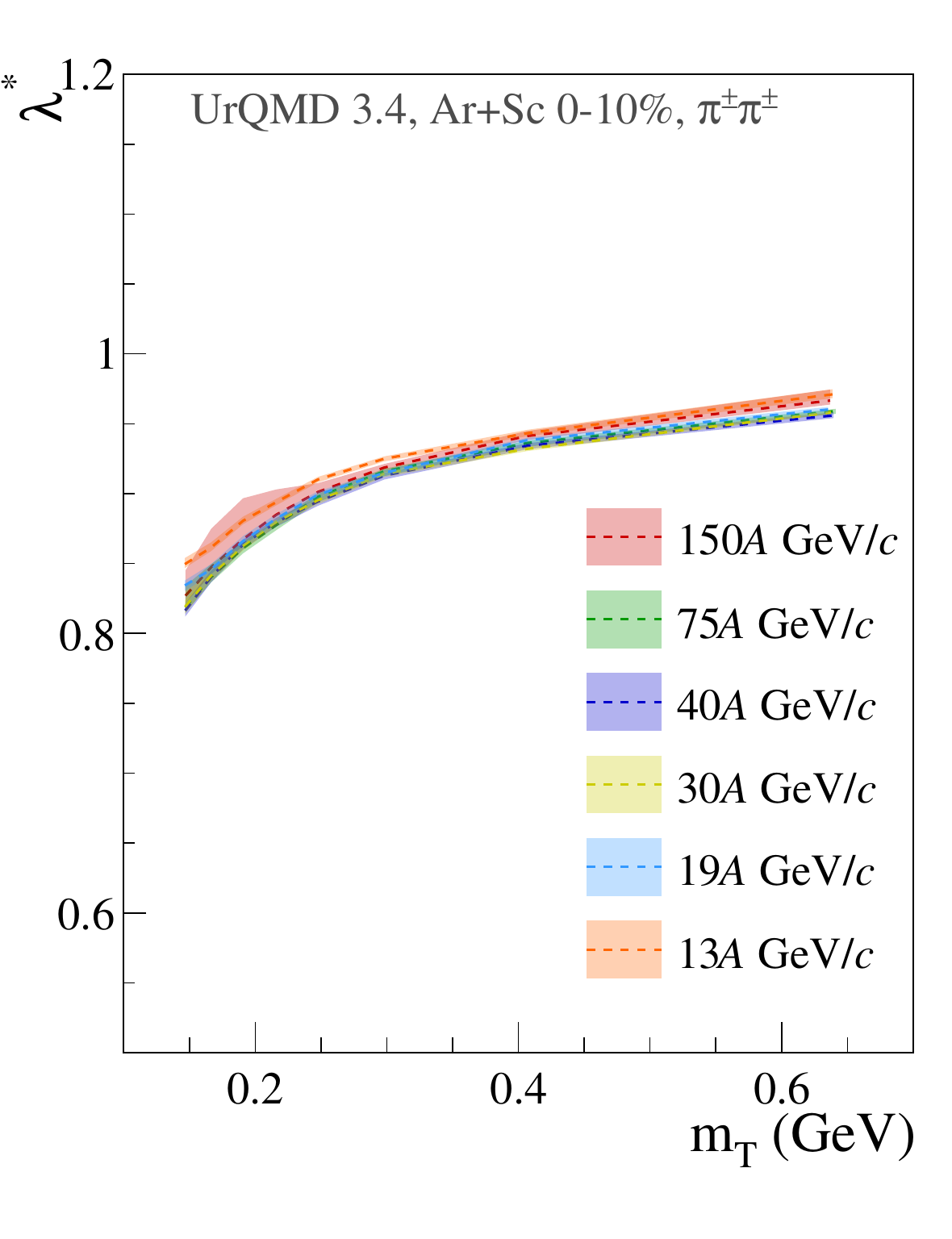}
     \caption{The intercept parameter $\lambda^*$, for 0--10\% central Ar+Sc at all beam momenta, as a function of transverse mass $m_\text{T}$. The given colored band shows systematic uncertainty.}
     \label{fig:lambda}
\end{figure}

Fig.~\ref{fig:lambda} shows the intercept parameter, or correlation strength parameter $\lambda^*$. We denote this parameter with an additional $^*$, as in UrQMD, even after modifying the code to include the decays of the $\eta$ meson, several weak decays are not included (those of kaons, $\Sigma$, $\Omega$, $\Xi$ baryons), and henceforth the pions stemming from the decays of these particles are missing from our simulation. These pions would all contribute to the halo (except for those from charged kaons, as these are produced so far out that they are removed from the pion sample in most experimental settings), thus our calculations result in a larger correlation strength. Therefore, direct comparison to $\lambda$ measurements is not possible. On the other hand, one can estimate the effect of the missing weak decays by calculating how many pions would have been produced had the above particles decayed, taking branching ratios also into account. We find that the ``true'' $\lambda$ would be about 92\%-93\% of the obtained value if these decays were incorporated, taking into account the yields and branching ratios to pions of the above particles (this fraction decreases very slowly with collision energy in the  13\textit{A}-150\textit{A} GeV/$c$ range). This correction aside, investigating the $\lambda^*$ versus $m_{\rm T}$ behavior, a small increase is found, while no clear collision energy dependence emerges.

\begin{figure*}
\centering
     \includegraphics[height=.47\textwidth, trim=4 4 21 4,clip]{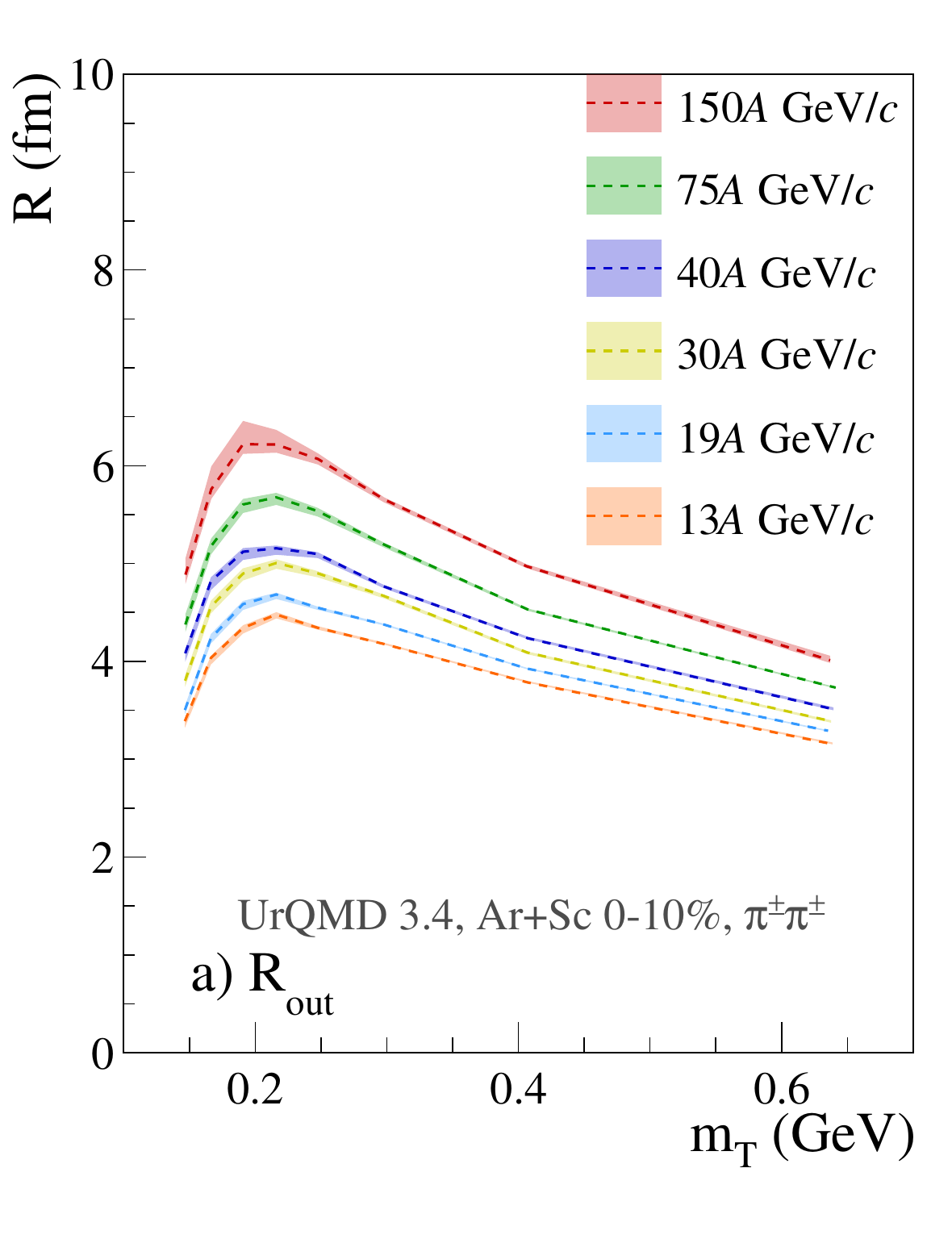}
     \hspace{-0.27cm}
     \includegraphics[height=.47\textwidth, trim=72 4 21 4,clip]{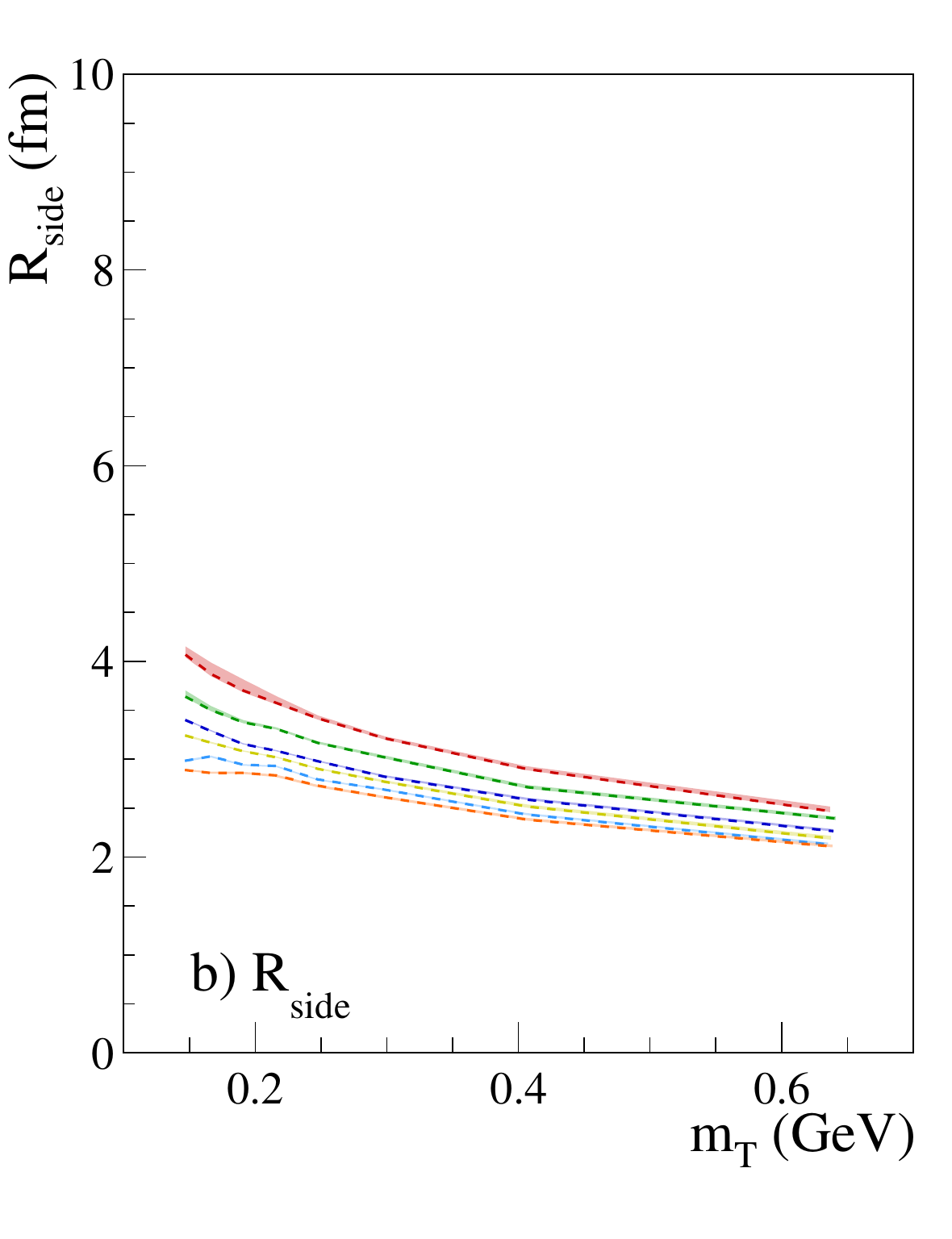}
     \hspace{-.27cm}
     \includegraphics[height=.47\textwidth, trim=72 4 21 4,clip]{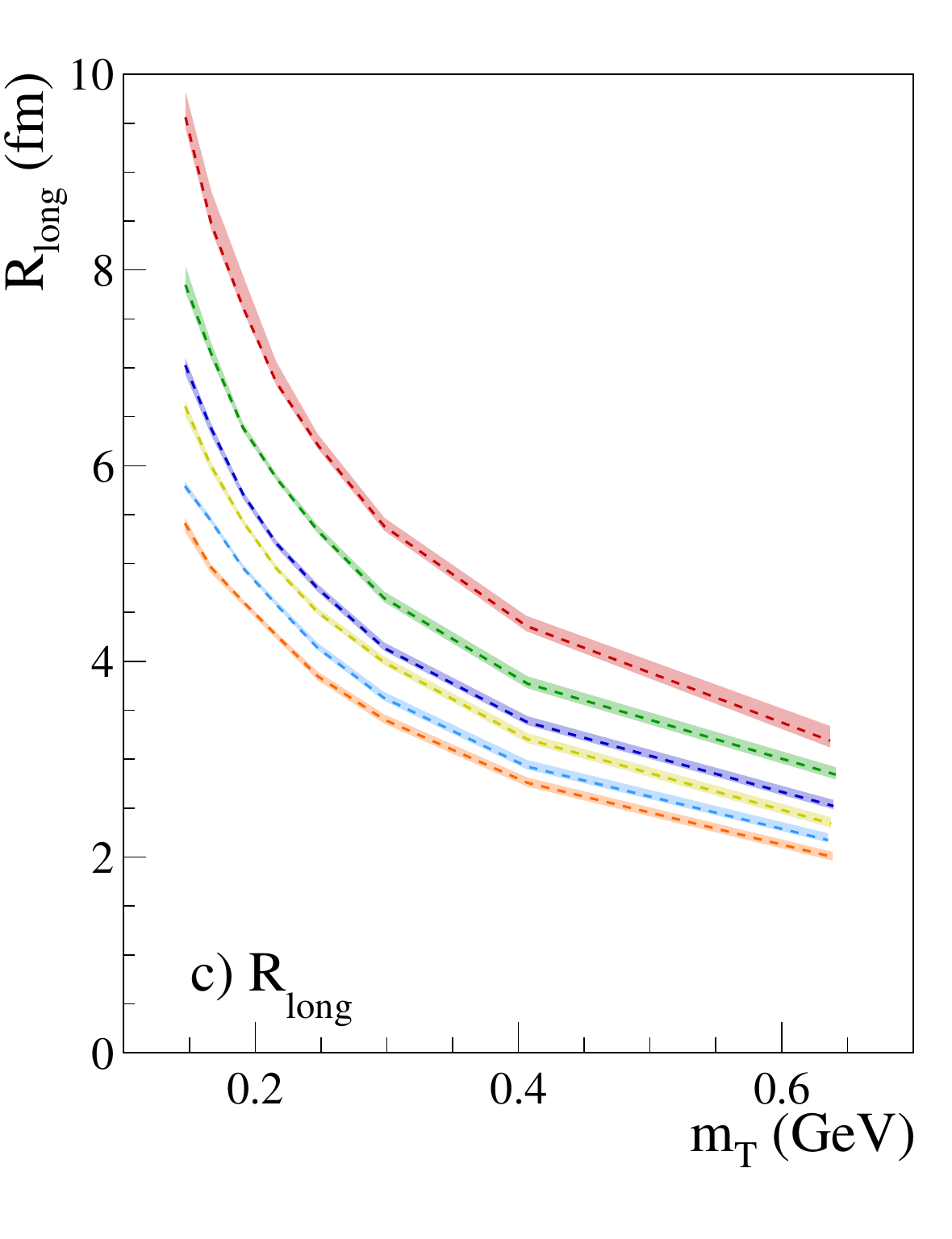}
     \caption{The Lévy-scale parameters $R_\text{out, side, long}$, for 0--10\% central Ar+Sc at all beam momenta, as a function of transverse mass $m_\text{T}$. The colored bands show the systematic uncertainty.}
     \label{fig:R}
\end{figure*}

\begin{figure*}
    \centering
     \includegraphics[width=.4\textwidth, trim=4 4 21 8,clip]{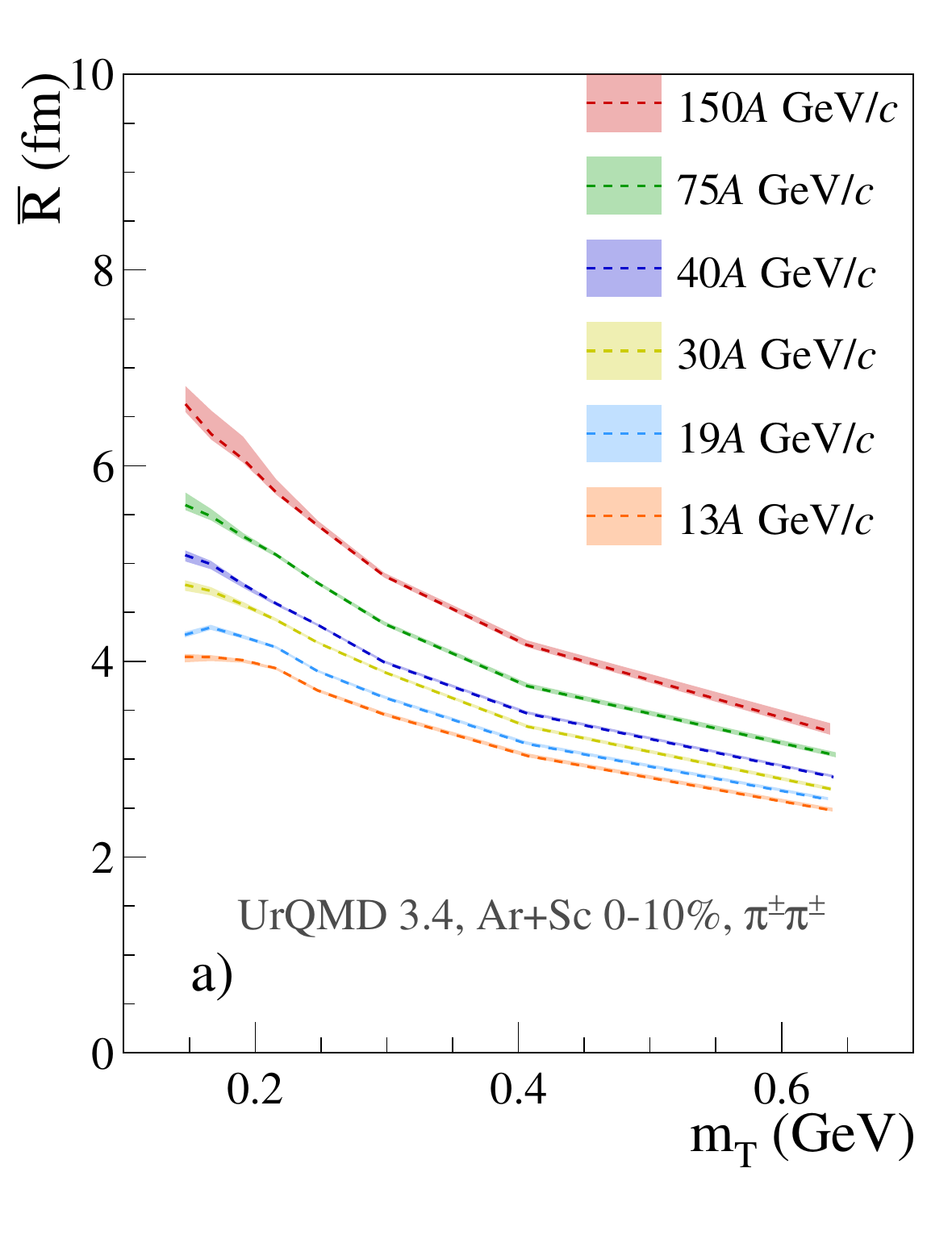}
     \includegraphics[width=.4\textwidth, trim=4 4 21 8,clip]{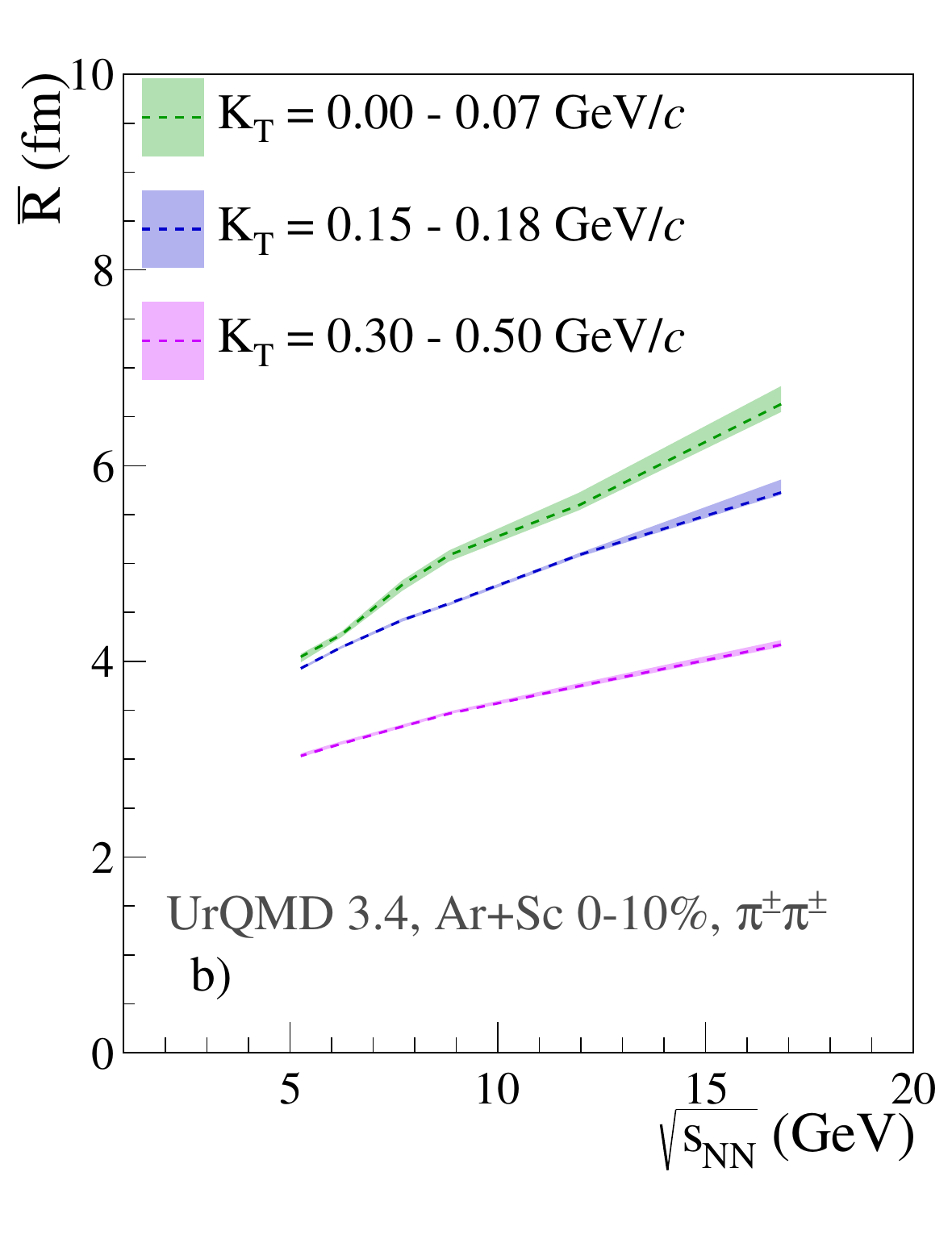}
     \caption{The averaged Lévy-scale parameter $\bar{R}$, for 0--10\% central Ar+Sc. For all transverse mass $m_\text{T}$ as a function of $m_\text{T}$ a) and for selected transverse mass $m_\text{T}$ intervals as a function of beam momentum b). The given colored band shows systematic uncertainty.}
     \label{fig:Rs}
\end{figure*}

Fig.~\ref{fig:R} shows the Lévy scale parameters $R_\text{out}$, $R_\text{side}$, and $R_\text{long}$. In all three directions, a clear decreasing trend with $m_\text{T}$ is observed, as traditionally attributed to an expanding source~\cite{PHENIX:2004yan} and collective flow, to some extent present in the microscopic UrQMD simulations, and observed in hydrodynamic models as well~\cite{Csorgo:1995bi}. On the other hand, several differences can be identified in terms of exact trends and values. Notably, in the ``out'' direction in the very low transverse mass regions, for all energies, a small maximum appears at around 0.2 GeV. There is also a clear $R_{\text{long}}>R_{\text{out}}>R_{\text{side}}$ ordering of the parameters at low $m_{\rm T}$. This changes at higher $m_{\text{T}}$ values: the $R_{\text{long}}$ becomes considerably smaller than $R_{\text{out}}$. This behavior is was also found in EPOS3~\cite{Kincses:2025izu} and EPOS4~\cite{Huang:2025edi} analyses, as well as experimental data~\cite{STAR:2014shf}. It is important furthermore to note that this HBT radii ordering has been seen to depend on the collision energy: $R_{\text{long}}$ increases with $\sqrt{s_{\text{NN}}}$  stronger than the other radii, due to its connection to the system lifetime~\cite{Makhlin:1987gm}, and becomes larger than the other radii for all $m_{\text{T}}$ values. This is apparent in both EPOS4~\cite{Huang:2025edi} as well as in experimental data~\cite{STAR:2014shf}. Both show in particular that in the $\sqrt{s_{\text{NN}}}<20$ GeV collision energy range, $R_{\text{out}}$ is larger than $R_{\text{long}}$. This behavior is also seen in our UrQMD results. This ordering is strongly connected to the medium properties and phase transition strength~\cite{STAR:2014shf}, and future three-dimensional measurements are required to distinguish between different scenarios and Equations of State.
To compare with one-dimensional femtoscopic analyses, it is also important to investigate the average $\bar{R} = \sqrt{\left(R_\text{out}^2+R_\text{side}^2+R_\text{long}^2\right)/3}$, shown as a function of $m_{\rm T}$ in Fig.~\ref{fig:Rs}a. Fig.~\ref{fig:Rs}b shows the collision energy dependence of the radii in three K$_\text{T}$ intervals. These exhibit a similar increase; however, their trends are slightly different, especially in the case of the lowest $K_{\rm T}$ bin, similarly to the case of $\langle \alpha\rangle$.

\section{Conclusion}\label{s:con}
We investigated three-dimensional two-pion spatial pair distance distributions in central $^{40}$Ar+$^{45}$Sc 0-10\% centrality collisions across SPS energies (13-150\textit{A} GeV/\textit{c}) using the UrQMD framework. The distributions, calculated on an event-by-event basis and aggregated for statistical convergence, are well-described by three-dimensional Lévy-stable distributions. The Lévy source parameters were investigated as a function of transverse mass and collision energy. The analysis revealed that the Lévy index $\alpha$ shows a moderate but clear increase with transverse mass, and a decrease with collision energy. We attribute the former to the relative weakening of the weight of strong decays in the source for larger momenta, while the latter can be due to a denser environment allowing for a more pronounced Lévy walk.  For the Lévy scale parameters, a decreasing trend with transverse mass and a clear increase with collision energy were found, in line with expectations based on transverse expansion. Furthermore, the correlation strength parameter $\lambda$ was also investigated, and no clear trends with energy were observed, but a small increase with transverse mass was found. We note that the extracted $\lambda$ values are expected to be approximately $92-93\%$ lower than the value quoted in this analysis, due to the omission of weak decays (e.g., $K^0_S, \Lambda$) in UrQMD. It is important to highlight that UrQMD includes only hadronic scattering, resonance decays, and coalescence, and lacks critical or hydrodynamic phenomena. Altogether, these results provide an important baseline for comparison with future experimental studies and with simulations of various collision systems and energies. 

\section*{Acknowledgments}
This research was funded by the NKFIH grants TKP2021-NKTA-64, K-146913, K-138136, and NKKP-152097. Barnab\'as P\'orfy furthermore acknowledges support of the DKOP-23 Doctoral Excellence Program of the Ministry for Culture and Innovation.

\bibliographystyle{elsarticle-num} 
\bibliography{HBT}

\appendix
\section{Ar+Sc UrQMD spectra}\label{a:spectra}

\begin{figure}[h!]
    \centering
     \includegraphics[width=.4\textwidth, trim=4 4 8 8,clip]{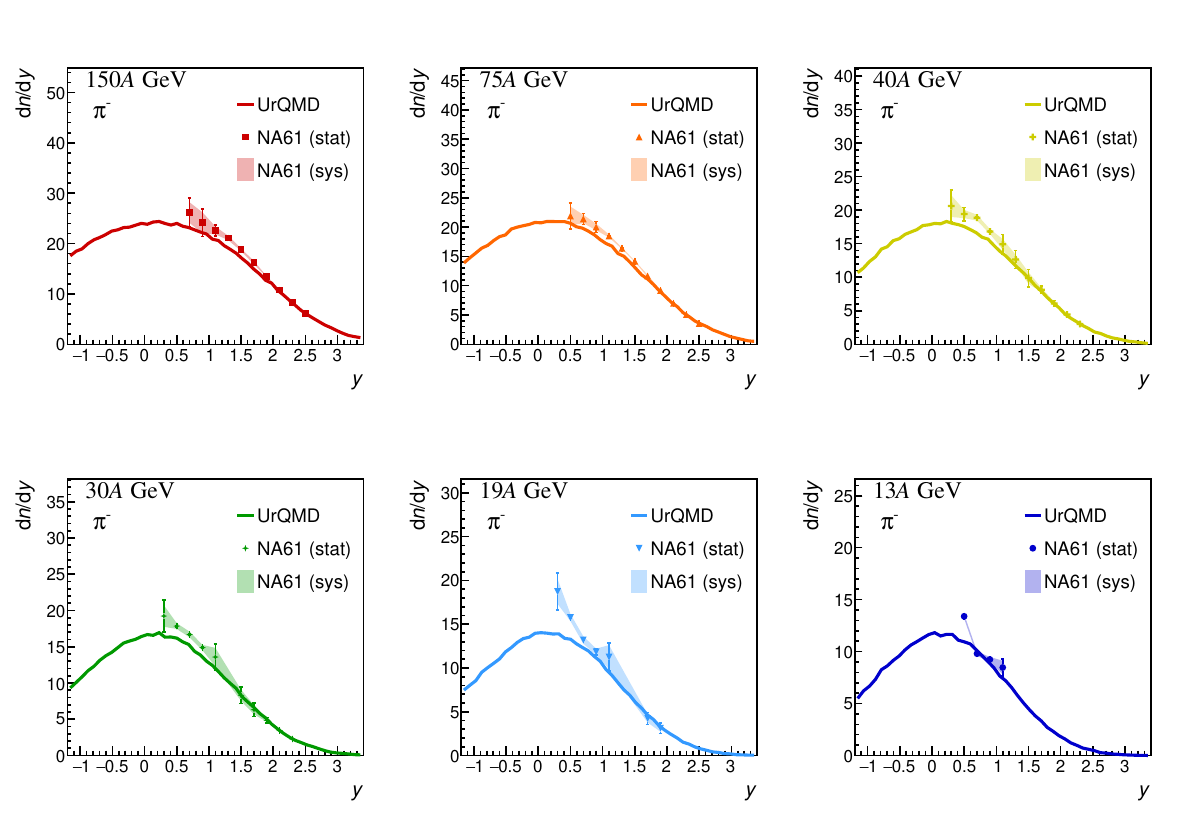}
     \includegraphics[width=.4\textwidth, trim=4 4 8 8,clip]{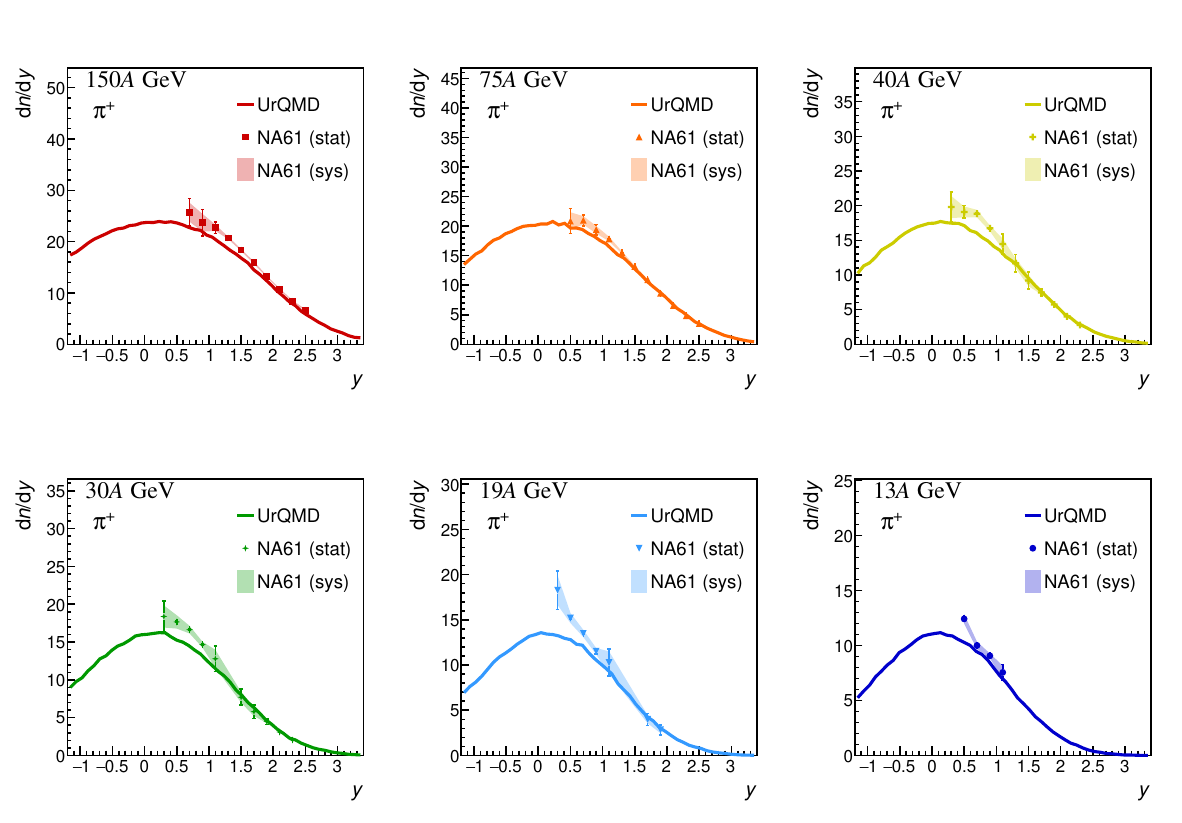}
     \includegraphics[width=.4\textwidth, trim=4 4 8 8,clip]{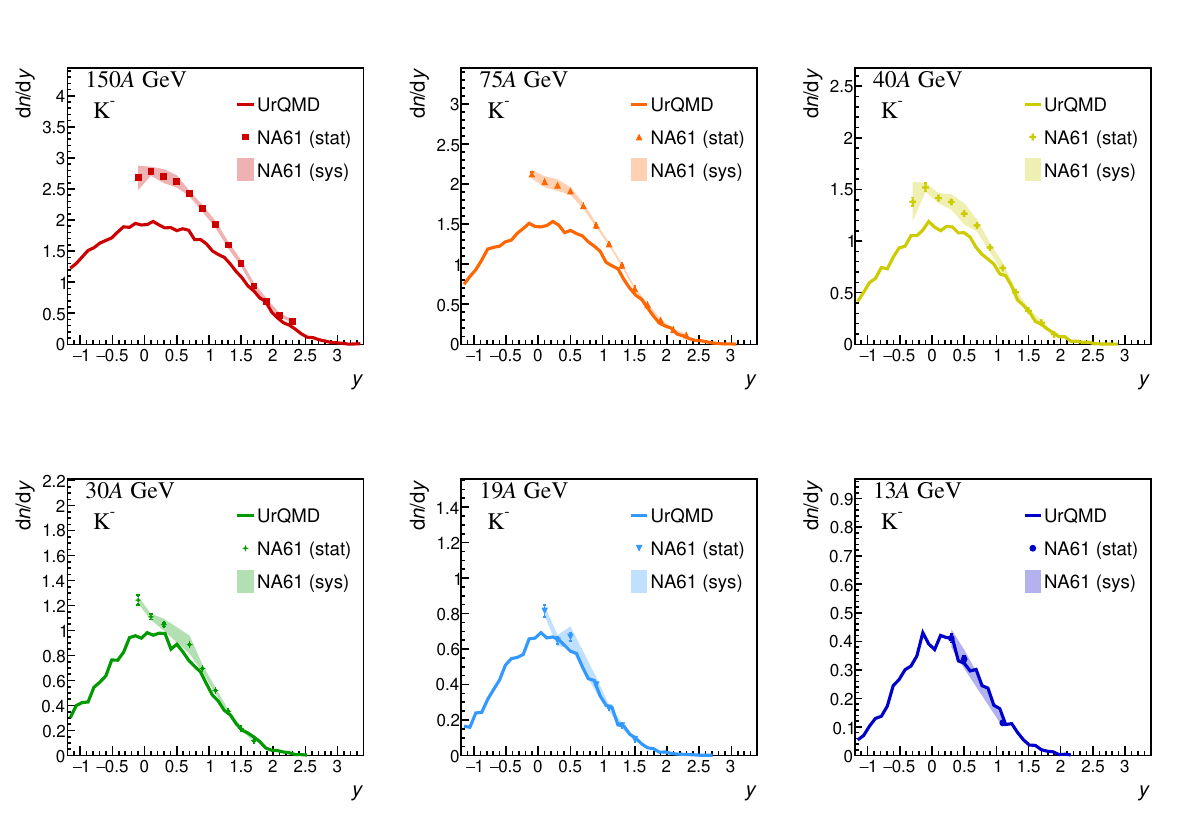}
     \includegraphics[width=.4\textwidth, trim=4 4 8 8,clip]{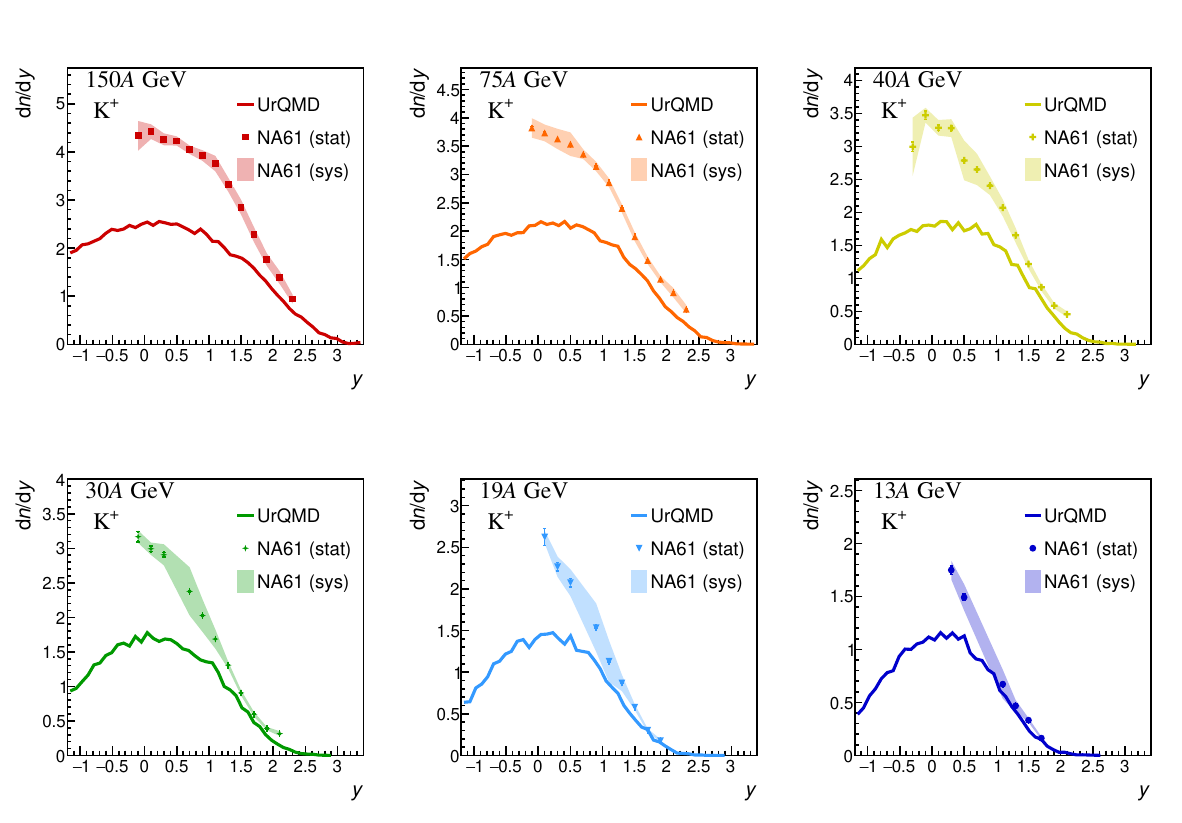}     
     \includegraphics[width=.4\textwidth, trim=4 4 8 8,clip]{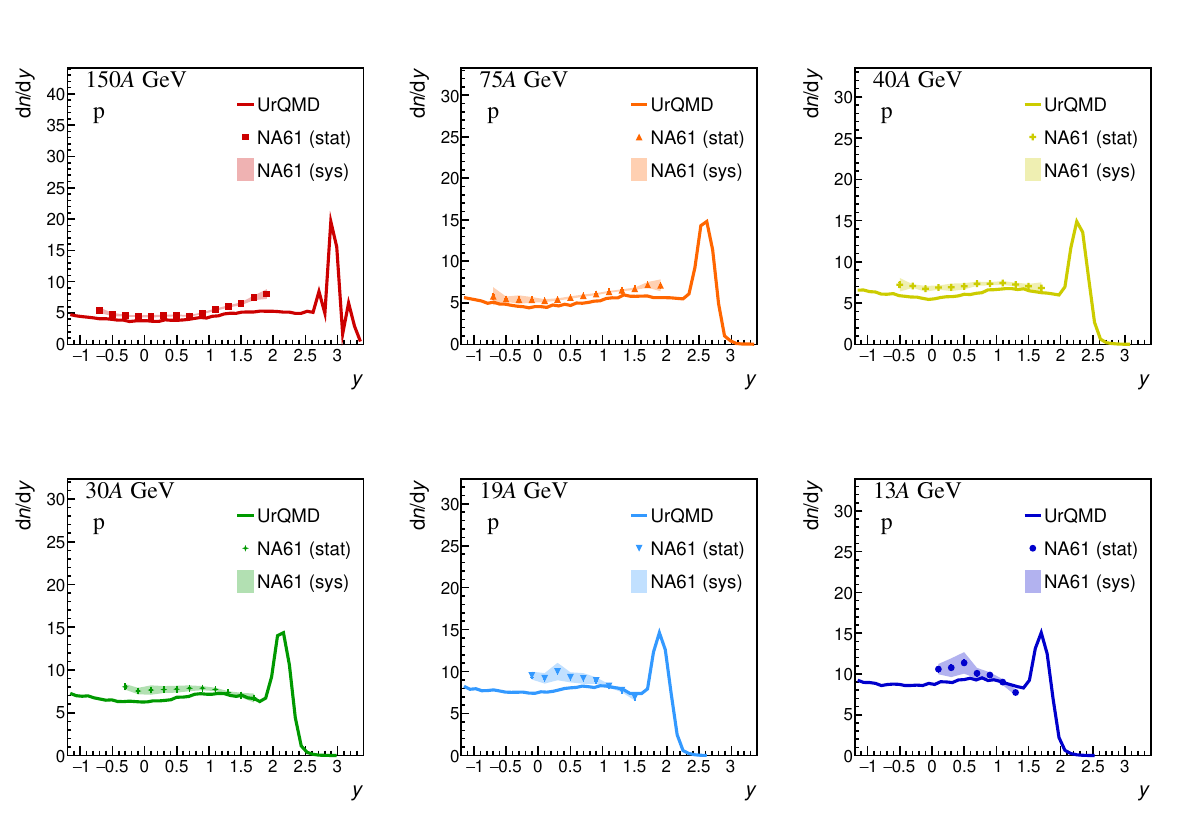}
     \includegraphics[width=.4\textwidth, trim=4 4 8 8,clip]{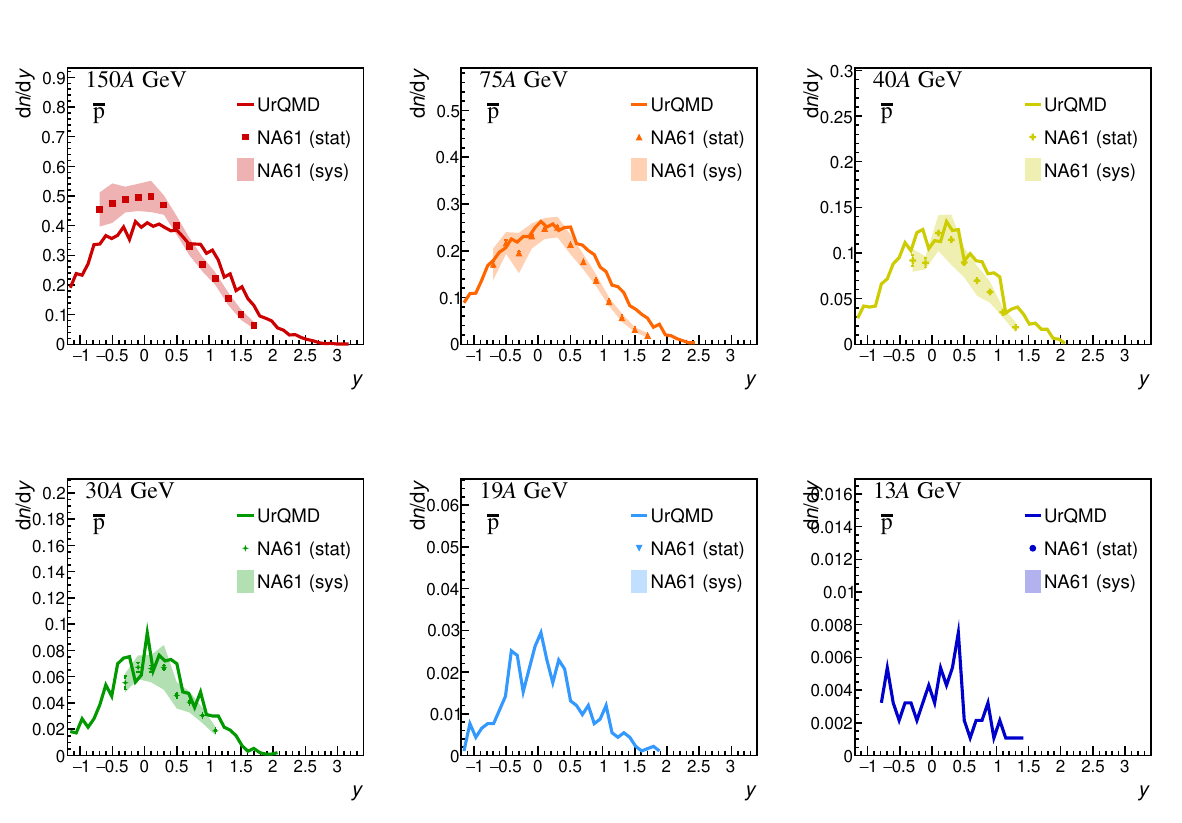}     
     \caption{Rapidity spectra of $\pi^\pm$, $K^\pm$, protons and antiprotons in 10\% most central, generated Ar+Sc collisions at 13\textit{A}, 19\textit{A}, 30\textit{A}, 40\textit{A}, 75\textit{A}, and 150\textit{A} GeV/\textit{c}, compared to published NA61/SHINE results.}
     \label{fig:dndy_all}
\end{figure}

\begin{figure}[h!]
    \centering
     \includegraphics[width=.4\textwidth, trim=4 4 8 8,clip]{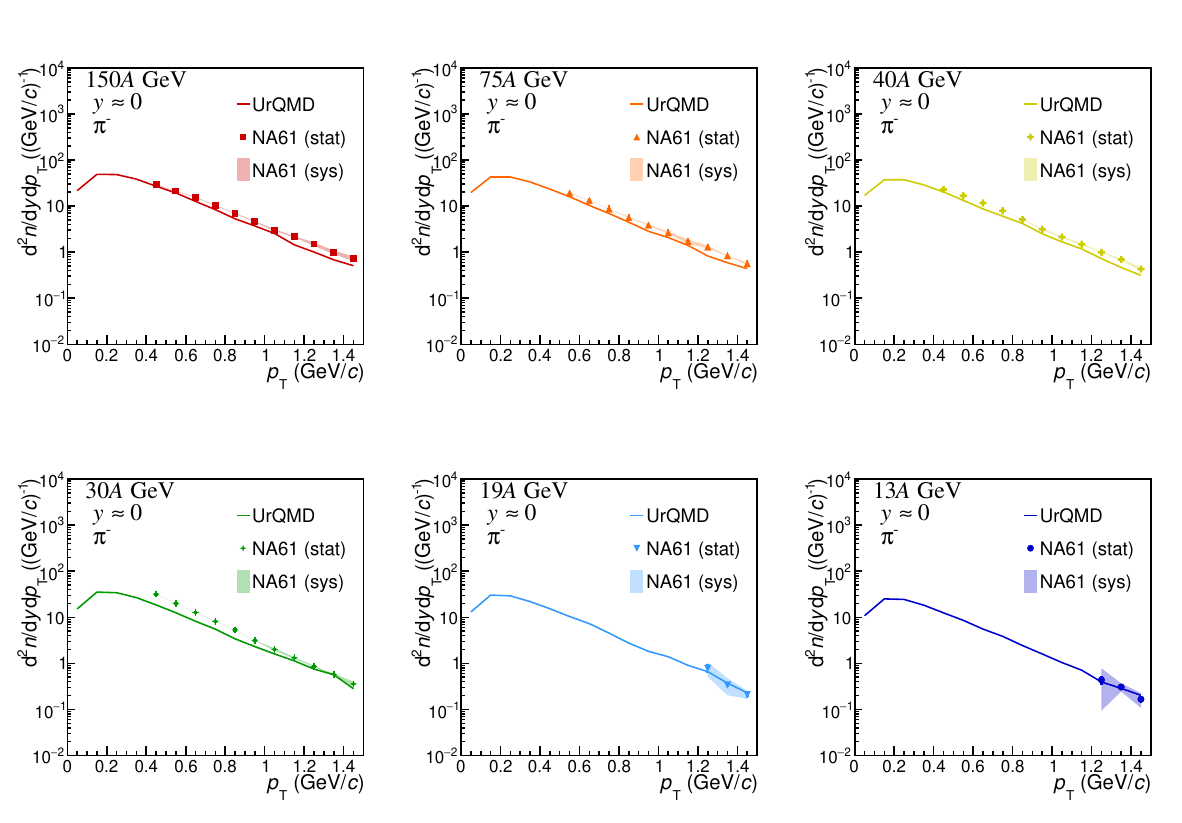}
     \includegraphics[width=.4\textwidth, trim=4 4 8 8,clip]{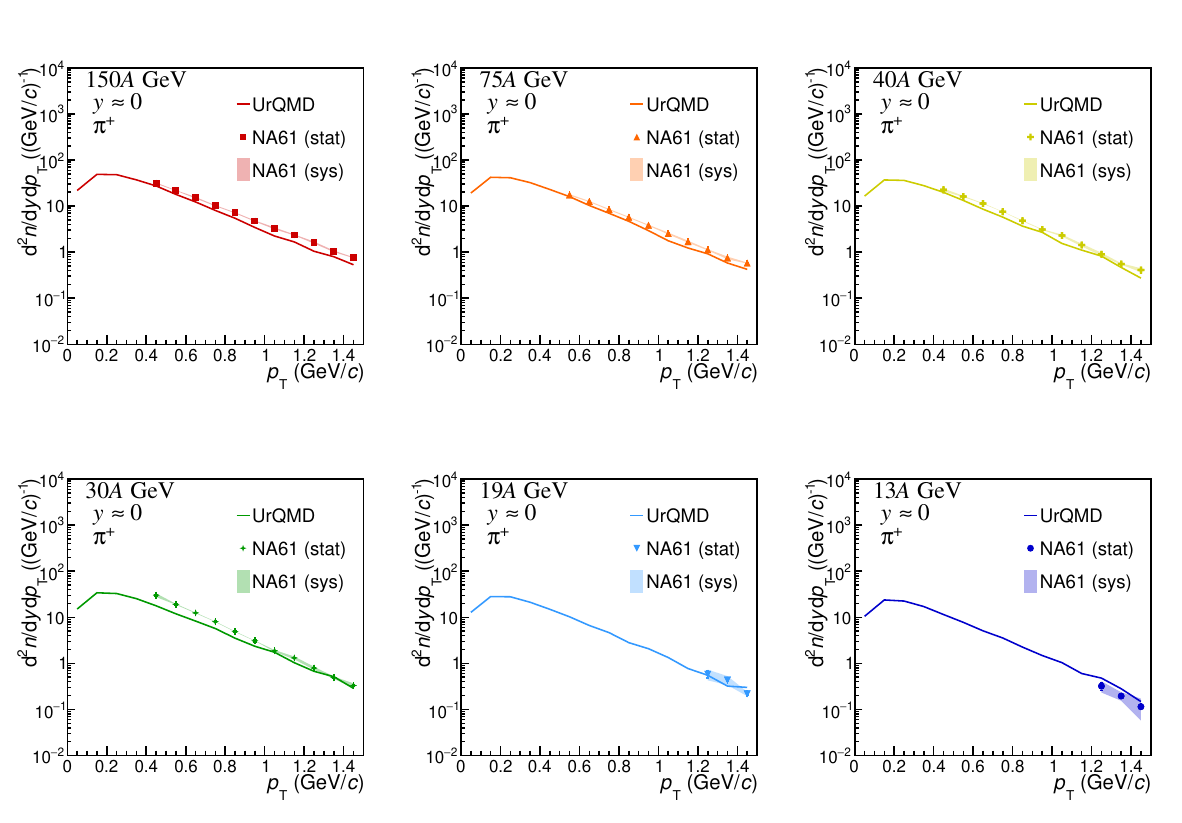}
     \includegraphics[width=.4\textwidth, trim=4 4 8 8,clip]{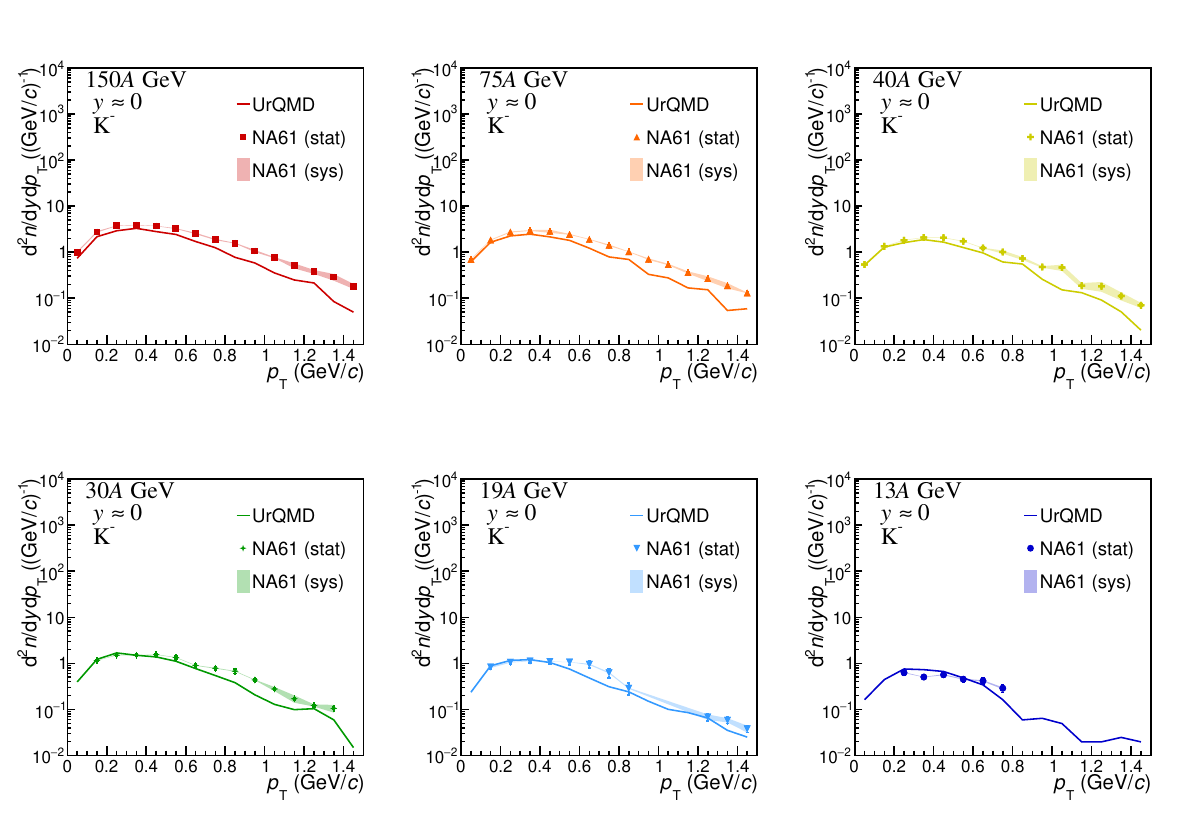}
     \includegraphics[width=.4\textwidth, trim=4 4 8 8,clip]{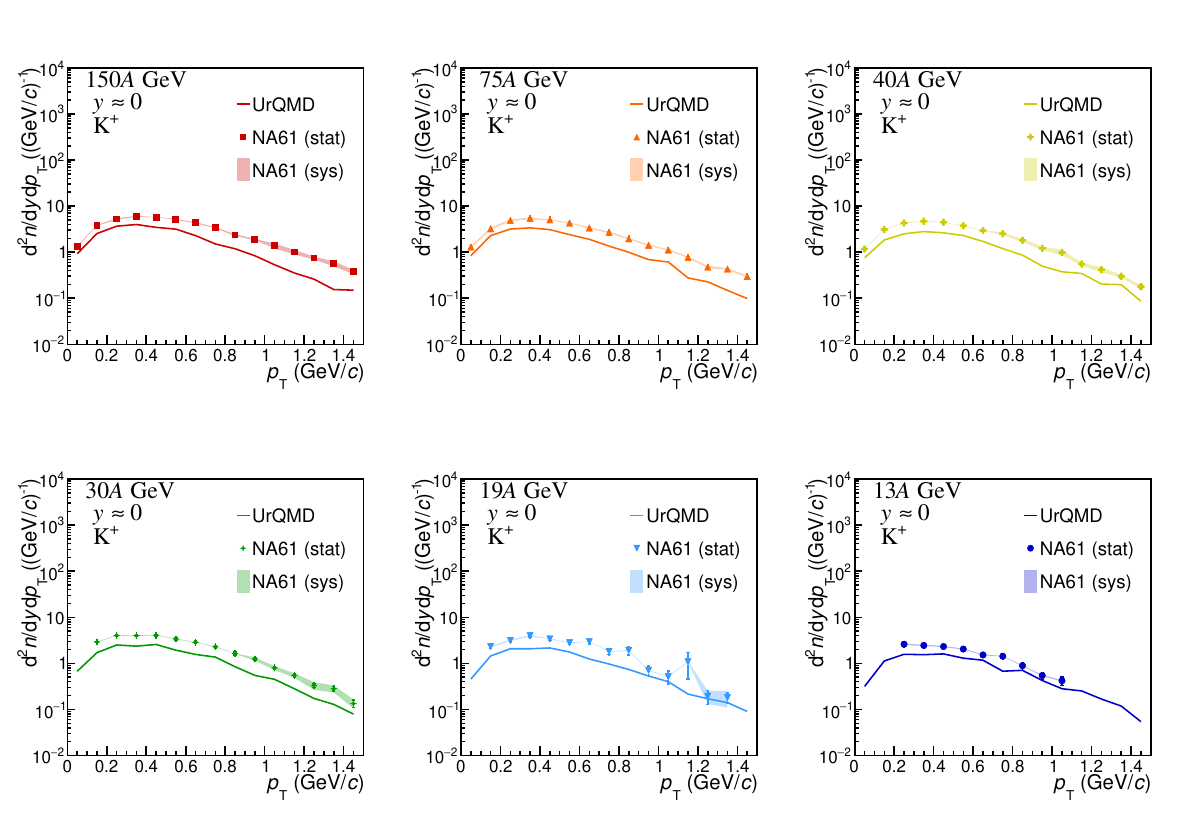}     
     \includegraphics[width=.4\textwidth, trim=4 4 8 8,clip]{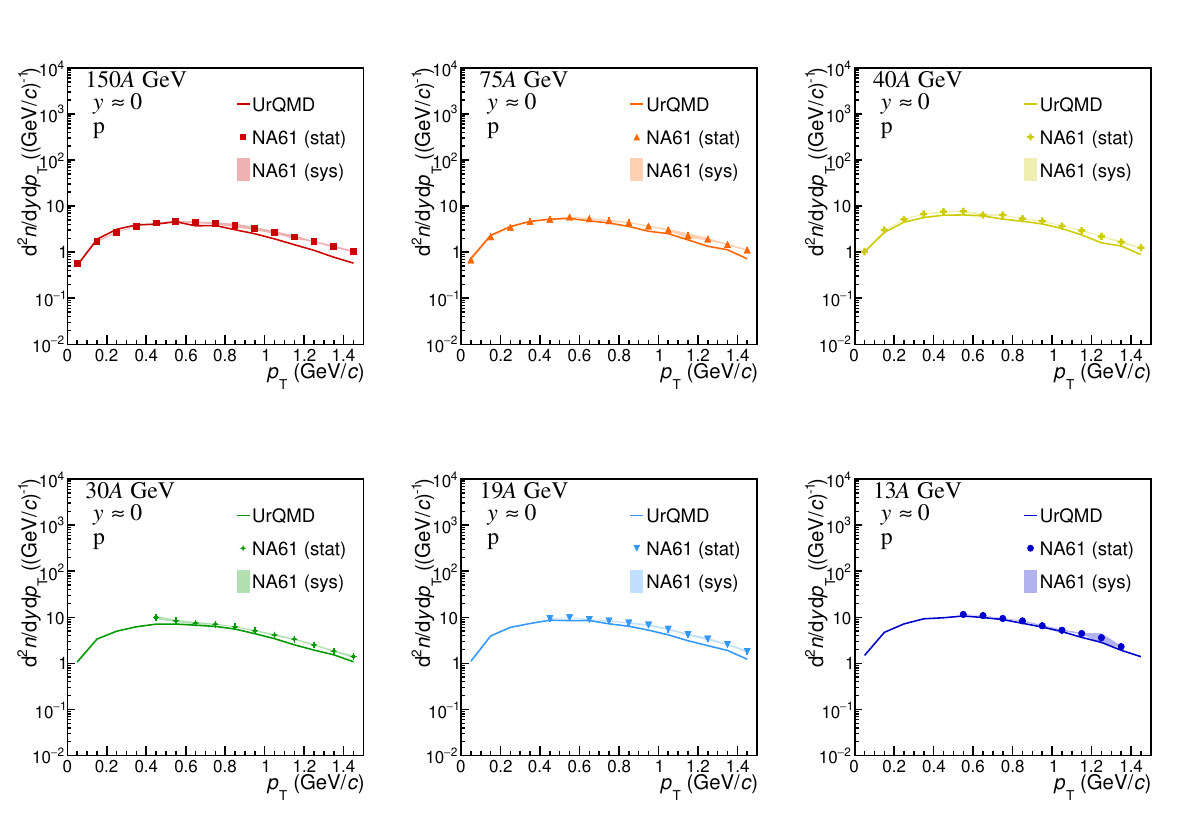}
     \includegraphics[width=.4\textwidth, trim=4 4 8 8,clip]{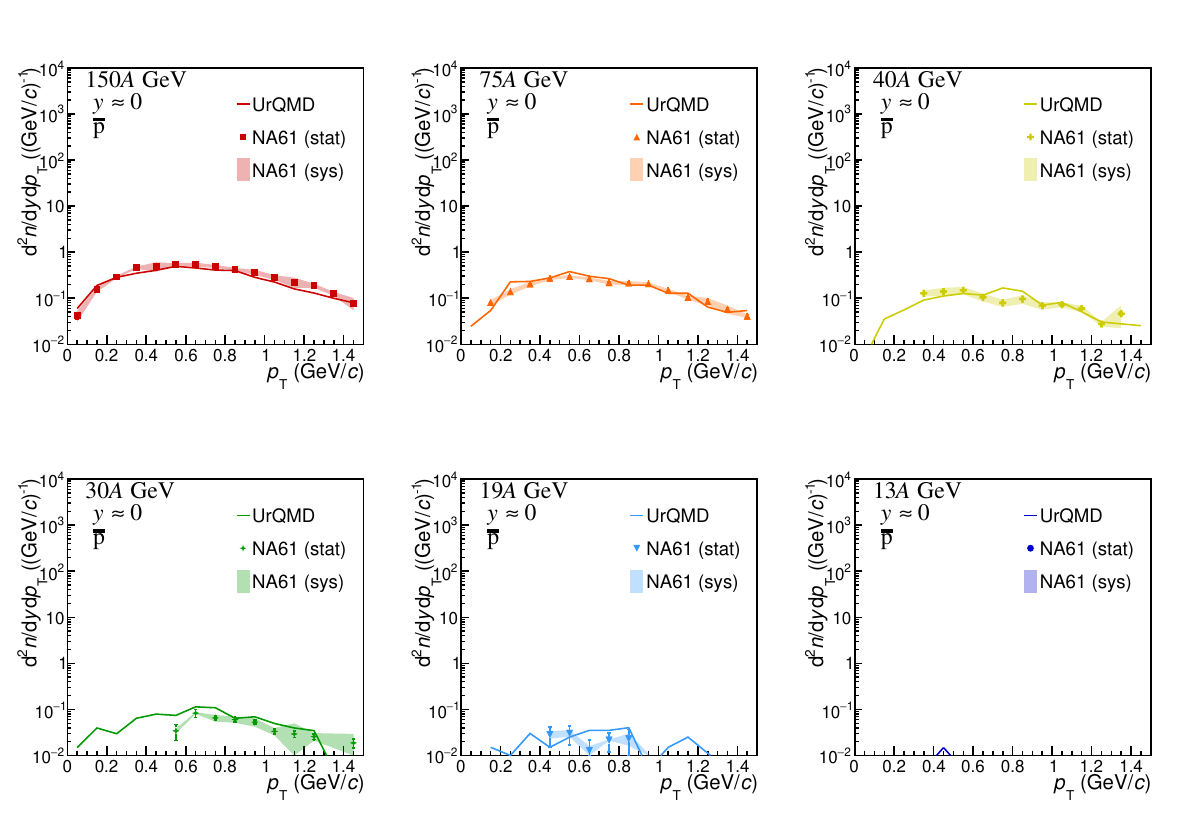}     
     \caption{Transverse momentum distribution of $\pi^\pm$, $K^\pm$, protons and antiprotons at midrapidity in 10\% most central, generated Ar+Sc collisions at 13\textit{A}, 19\textit{A}, 30\textit{A}, 40\textit{A}, 75\textit{A}, and 150\textit{A} GeV/\textit{c}, compared to published NA61/SHINE results.}
     \label{fig:d2ndpt_all}
\end{figure}

\begin{figure}[h!]
    \centering
     \includegraphics[width=.4\textwidth, trim=0 4 8 8,clip]{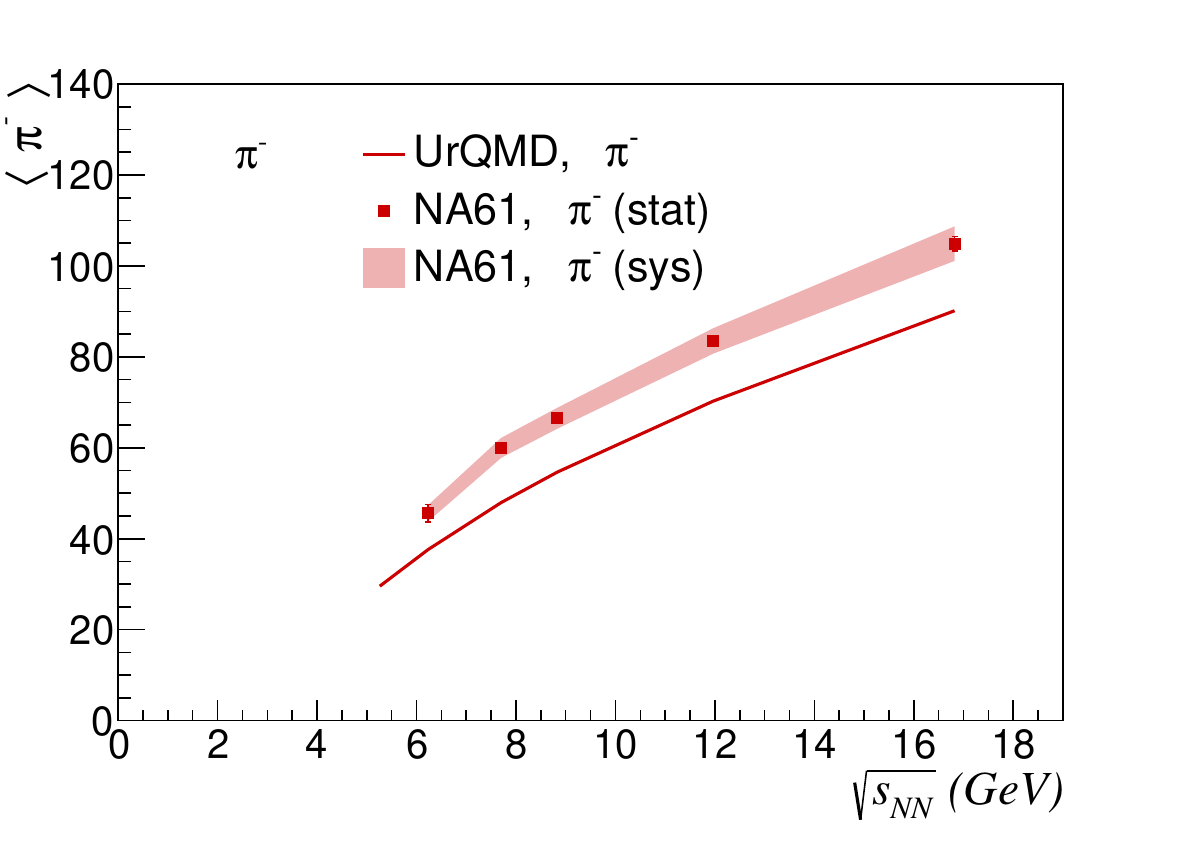}
     \includegraphics[width=.4\textwidth, trim=0 4 8 8,clip]{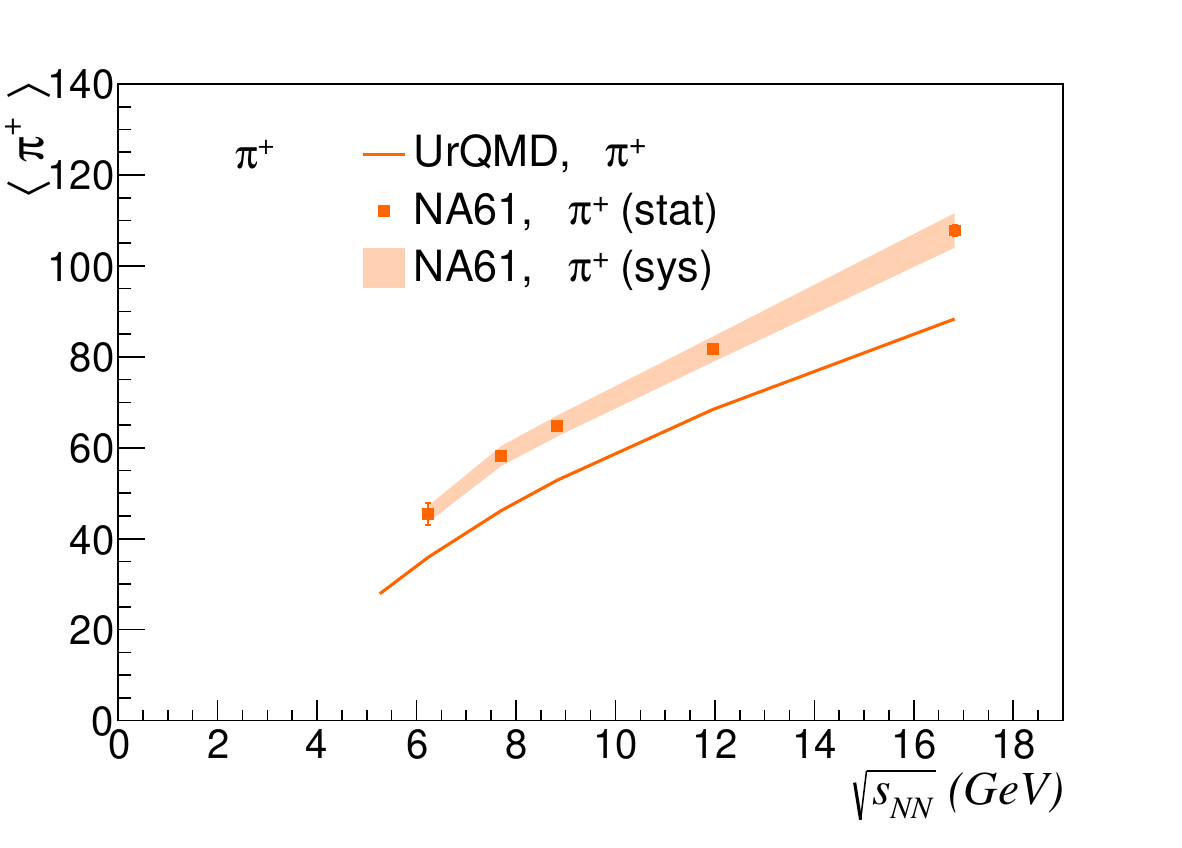}
     \includegraphics[width=.4\textwidth, trim=0 4 8 8,clip]{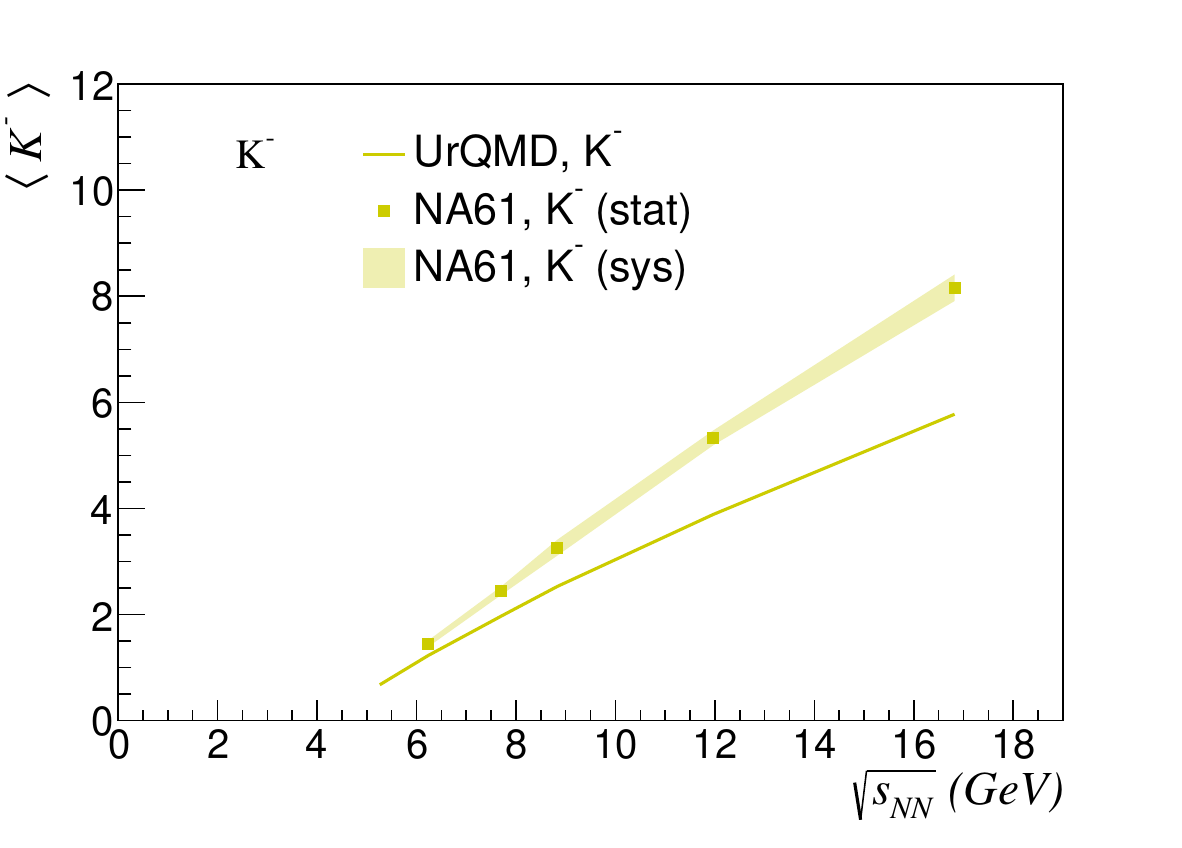}
     \includegraphics[width=.4\textwidth, trim=0 4 8 8,clip]{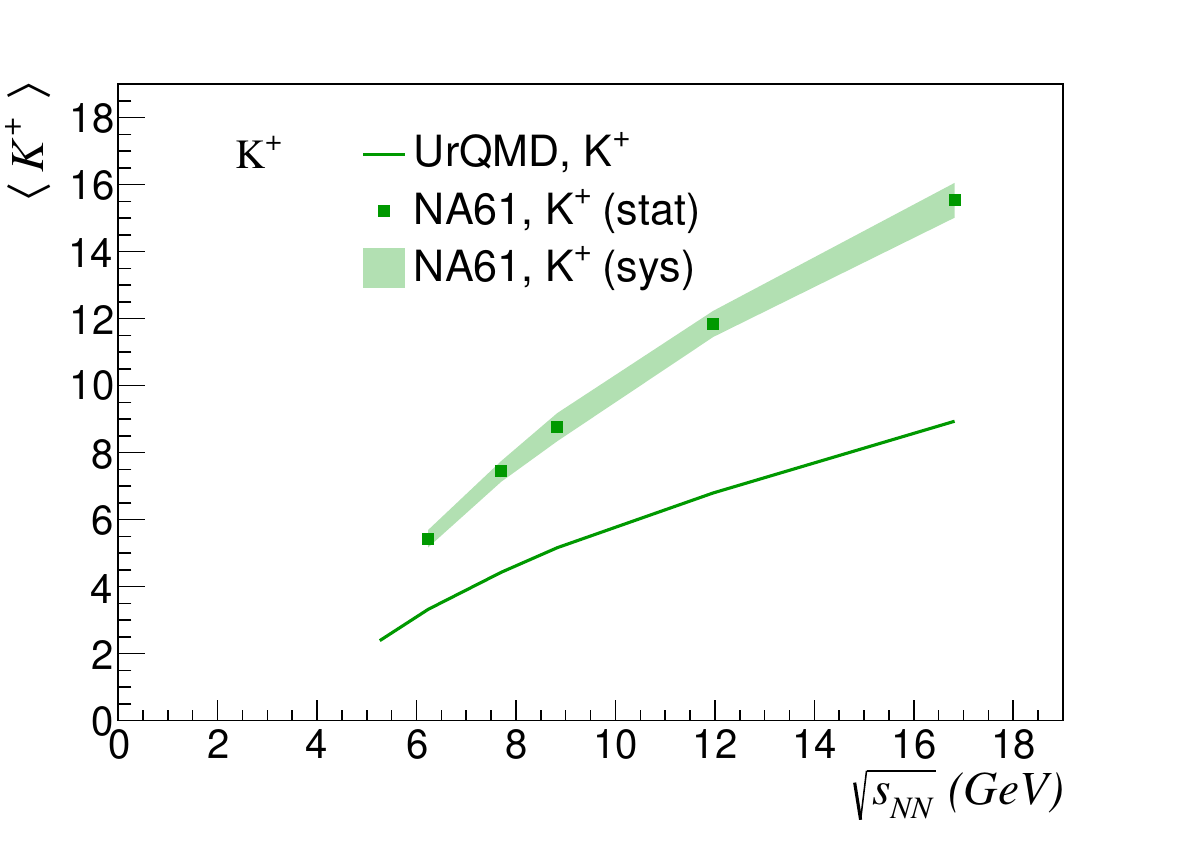}      
     \caption{Mean multiplicities of $\pi^\pm$ and $K^\pm$ in 10\% most central, generated Ar+Sc collisions at 13\textit{A}, 19\textit{A}, 30\textit{A}, 40\textit{A}, 75\textit{A}, and 150\textit{A} GeV/\textit{c}.}
     \label{fig:mean_mult_all}
\end{figure}

\end{document}